\documentclass[12pt]{article}
\pdfoutput=1
\usepackage[top=30truemm,bottom=30truemm,left=25truemm,right=25truemm]{geometry}
\usepackage{authblk}
\usepackage{amsthm}
\usepackage{amsmath,amssymb,bm,mathtools}
\usepackage{cite}
\usepackage[]{hyperref}
\hypersetup{colorlinks,linkcolor={blue},citecolor={blue},urlcolor={blue}}  
\usepackage[titletoc]{appendix}
\usepackage{url}
\usepackage{graphicx,color}
\usepackage{braket}
\usepackage{qcircuit}
\usepackage{here}
\usepackage{mathrsfs}
\usepackage{subcaption}
\usepackage{geometry}
\usepackage{titlesec}

\newtheorem*{definition*}{Definition}
\newtheorem*{assumption*}{Assumption}

\def\Tr{{\mathrm{Tr}}}

\def\sinh{{\mathrm{sinh}}}
\def\cosh{{\mathrm{cosh}}}
\def\tanh{{\mathrm{tanh}}}

\def\CM{{\cal M}}

\def\be{\begin{equation}}
\def\ee{\end{equation}}
\def\ba{\begin{eqnarray}}
\def\ea{\end{eqnarray}}

\numberwithin{equation}{section}

\usepackage{setspace} 

\AtBeginDocument{%
  {\footnotesize\global\footnotesep=0.8\baselineskip}%
}

\definecolor{cardinal}{rgb}{0.6,0,0}
\definecolor{darkgreen}{rgb}{0,0.4,0}
\definecolor{golden}{rgb}{0.92, 0.7, 0}
\definecolor{midnight}{rgb}{0, 0, 0.5}
\definecolor{darkblue}{rgb}{0, 0, 0.7}
\definecolor{purple}{rgb}{0.5, 0, 0.5}

\begin{document}

\begin{titlepage}
\thispagestyle{empty}

\begin{flushright}
\end{flushright}

\bigskip

\begin{center}
\noindent{\bf \Large Observers and Timekeepers: From the Page-Wootters }
\vskip 0.2cm
\noindent{\bf \Large  Mechanism to the Gravitational Path Integral}
\vskip 1cm


{\bf Zixia Wei }
\vspace{0.5cm}


{\it
Society of Fellows,
Harvard University, Cambridge, MA 02138, USA
}\\[1.5mm]

\vspace{0.3cm}

{\tt  zixiawei@fas.harvard.edu\\}

\medskip

\end{center}

\begin{abstract}

Quantum gravity in a closed universe faces two {\it a priori} distinct yet seemingly related issues: the problem of time and the fact that its Hilbert space dimension is one. Both have been argued to be resolvable by formulating physics relative to an observer. Using a simple gravitational path integral model, we explain that the two issues arise from two distinct non-perturbative effects: the former from summing over metrics and the latter from summing over topologies. We then revisit the Page-Wootters mechanism, one of the earliest frameworks for formulating quantum mechanics relative to an observer, see how it applies to both issues, and introduce some new ingredients. In particular, we emphasize a hierarchy between an observer and a timekeeper. An observer is a subsystem of the universe whose specification results in a nontrivial Hilbert space, while a timekeeper is an observer with a specified history that can be used as a reference for the time of the environment and experiences a nontrivial time evolution. Finally, we propose a method for incorporating observers and timekeepers into the gravitational path integral and show that implementing a timekeeper in this way furnishes an observer-dependent generalization of holography.

\end{abstract}
\medskip

\end{titlepage}

\newpage
\setcounter{page}{1}
\tableofcontents

\newpage

\section{Introduction}
At the very early stage of attempting to quantize gravity, the canonical formalism was used and a conceptual problem was encountered. In this approach, the Hamiltonian constraint is uplifted to an operator equation called the Wheeler-DeWitt (WDW) equation \cite{DeWitt67,Wheeler68}, which may be formally written as $H\ket{\Psi}_{\rm uni} = 0$ in a spatially closed universe. In WDW equation, the wave function of the universe $\ket{\Psi}_{\rm uni}$ is an eigenfunction of the Hamiltonian $H$, indicating that one would see a universe without time dependence if one were to interpret the WDW equation as a Schr\"odinger equation, which contradicts our daily experience. This issue is called the problem of time, and it had sat at the core of the study in quantization of spacetime geometry before late 1990s \cite{Kuchar91,Isham92}. 

The situation has been changed, however, after the discovery of the AdS/CFT correspondence \cite{Maldacena97,GKP98,Witten98}, which has attracted much attention to the study of quantum gravity in anti-de Sitter (AdS) spacetime. The AdS spacetime is an open universe, i.e. it has a timelike asymptotic boundary, on which the Dirichlet boundary condition is imposed and the metric is frozen. The Arnowitt–Deser–Misner (ADM) Hamiltonian now include a boundary charge term \cite{CGPR21,AKW22,Witten22}, which generates a time translation along the asymptotic boundary and hence does not have the problem of time. This nice feature makes the dynamics in AdS easier to study, but at the same time, it makes the problem of time overlooked in this context. 

Recently, significant progress has been made in closed universe quantum gravity using gravitational path integral \cite{MM20,UWZ24,UZ24}. In particular, it has been found that the Hilbert space of a closed universe is one-dimensional if one takes the contribution from wormhole configurations into account. This again, seems to contradict our daily observation, where we explore a diverse world, and see more than one states. For convenience, we call this feature of the closed universe the problem of dimension, in contrast to the problem of time. 

Mathematically, the problem of time and the problem of dimension are two independent issues. Given a Hilbert space, the problem of time starts from identifying an operator as the Hamiltonian, and impose $H\ket{\Psi}_{\rm uni} = 0$ as a constraint. The solutions of the constraining equation do not have an explicit time evolution under the Hamiltonian chosen above, but it does not prevent the solution subspace from being more than one-dimensional. On the other hand, the notion of a Hamiltonian does not necessarily show up in the argument of the problem of dimension, making the time meaningless. 

Nonetheless, the two issues possess many similarities from physical points of view. First of all, a closed universe is a self-contained system and does not interact with anything else. As a result, the only way to change it is through its unitary time evolution \cite{PSW05}. If there is no time evolution (the problem of time), the universe stays in one state forever (the problem of dimension). The other way around, if there is only one state in the Hilbert space, it is impossible to construct a nontrivial unitary operator other than a pure phase, indicating the absence of a nontrivial time evolution. 

Given the distinctions and similarities of the two issues, it should be instructive to learn how they arise from the same quantization procedure. 
This is the first purpose of this paper. 
In section \ref{sec:two_issues}, we study a model of 1D gravity and consider its path integral quantization. The advantage of this 1D model is that its Hilbert space can be constructed explicitly and tracked transparently in each step of the analysis. Also, despite its simpleness, it is rich enough to contrast the origins of the two issues. 
We will see that, while both of them arise from summing over geometries in the gravitational path integral, the problem of time originates from summing over metrics (more precisely, the lapse function) and the problem of dimension originates from summing over topologies. The same results follow in higher dimensions, although writing down the Hilbert explicitly is challenging. 
Many of the contents presented in section \ref{sec:two_issues} are a reformulation of some old results \cite{Halliwell88,HH90,Marolf00} and recent results \cite{MM20,UWZ24,UZ24} in a unified manner, while some new results will also be presented. 

In one word, it will become clear that the problem of time (dimension) arises in the gravitational path integral quantization because the metrics (topologies) are summed over {\it too nicely}. It follows immediately that if one sums over geometries in an unbalanced way, say dismiss some geometries or sum over them according to an unbalanced weight, then a nontrivial time evolution and nontrivial Hilbert space dimension will be retrieved.  
The question is then to find a physically sensible mechanism to make this happen. 

Leaving aside the gravitational path integral (or even gravity), the problem of time has been tackled by Page and Wootters (PW) in a purely quantum mechanical setup \cite{PW83,Wootters84}. 
The essence of the PW mechanism is to realize that, while the standard quantum mechanics is formulated for an external observer sitting outside the observed system, an observer is also a part of the universe when it is closed. Therefore, one should identify a proper subsystem of the closed universe as an observer\footnote{The observer here is called a clock in the original work \cite{PW83}.}, and then construct a variation of quantum mechanics relative to the chosen observer. 
The PW mechanism takes the existence of the Hilbert space and the Hamiltonian constraint as the starting point and proposed a way to retrieve a nontrivial time evolution
relative to the observer. 
In this formalism, all the possible histories\footnote{We use ``history" to mean a trajectory of time evolution in this paper.} are encoded in the entanglement structure between the observer and their complement, which is identified as the observed environment. Once a history of the observer is chosen among all possibilities, the PW mechanism allows us to read out the corresponding history of the environment relative to the observer. We will revisit the PW mechanism in section \ref{sec:PW}, and emphasize two technically straightforward but conceptually important points along the way. One is that the PW mechanism automatically solve the problem of dimension in a purely quantum mechanical setup, although it is not usually formulated in that way. 
The other is that it turns out to be crucial to distinguish an observer with a specified history and that without one for later uses. We will call an observer with a specified history a timekeeper, in the sense that they can use their own history to refer time. By contrast, when we say an observer, we mean one without a specified history unless otherwise stated. 

The second purpose of this paper is to 
propose a physically sensible formulation of an observer and a timekeeper in the gravitational path integral as an analogue of the PW mechanism. 
In fact, different proposals for incorporating an observer in the gravitational path integral has been made very recently \cite{BNU23,HUZ25,AAIL25} to cope with the problem of dimension. 
See also \cite{BKU25,Chen25,NU25} for related developments. 
As we have mentioned above, since any mechanism which makes the topologies summed over in an unbalanced way will generically result in a nontrivial Hilbert space dimension, it is not surprising that there can be multiple of them. 
On the other hand, the discussion on the problem of time, the implementation of a timekeeper, and the explicit comparison with the PW mechanism were not presented in these literature.

We aim to give a formulation incorporating both the observer and the timekeeper with a clear hierarchy between them, similar to that in the PW mechanism.
More specifically, an observer will be implemented as an extended object, who themselves are matter and a subsystem of the universe. We will see that specifying such an observer puts restrictions on summing over topologies and hence results in a nontrivial Hilbert space, similar to that in \cite{BNU23,HUZ25,AAIL25}. A timekeeper will be formulated as an observer with a frozen intrinsic geometry, specifying which puts restrictions on summing over spacetime metrics and hence results in a nontrivial time evolution. Such path integrals may be called ``gravitational path integral relative to an observer/timekeeper". Moreover, and interestingly, we will show that gravitational path integrals relative to a timekeeper includes the Gubser-Klebanov-Polyakov-Witten (GKP-Witten) \cite{GKP98,Witten98} formulation of the AdS/CFT correspondence \cite{Maldacena97} as a special case, and they furnish an observer-dependent generalization of the holographic principle \cite{tHooft93,Susskind94}. 

A general formulation will be presented in section \ref{sec:observer_timekeeper}, accompanied by its implementation in the 1D gravity example. Since the 1D gravity with a timekeeper turns out to be too simple, we construct a 3D gravity model incorporating all the features mentioned above in section \ref{sec:3D_gravity}. This model uses AdS with gravitating (i.e. not frozen) boundaries to realize a closed universe, and matters in it as observers and timekeepers. This feature makes the relation between the gravitational path integral relative to an observer or a timekeeper and well-known holographic correspondences transparent. 
We will also discuss how to incorporate multiple observers/timekeepers in this formulation, with which the observer dependence of gravitational path integral manifests. 

Along the way of constructing the gravitational path integral relative to an observer or a timekeeper, its qualitative similarities to the PW mechanism will be made clear. One may wonder how this can be useful in the future study of quantum gravity, given the non-gravitational nature of the PW mechanism and the similarities being only qualitative. In fact, the PW mechanism plays an essential role in another actively studied field, the quantum reference frame \cite{AS67,BRS07,GS08,GCB17,VHGC18,HSL19}.
Studies on the quantum reference frame aim to formulate quantum mechanics in a fully relational way without introducing any absolute reference such as an ideal time direction. Its application to perturbative quantum gravity has also been discussed (See \cite{GHK22,DVEHK24,Kirklin24} for some recent work). 
Besides, it has been recently argued \cite{DVEHK24} that the PW mechanism in quantum reference frames has a profound relation with the von Neumann algebra approach \cite{LL21-1,LL21-2,Witten21,CLPW22,Witten23} of quantum gravity. Moreover, there are many interesting and relatively well-understood features and technics in the quantum reference frame, such as the observer dependence of the physical observations and the switching between the observers \cite{GCB17,VHGC18}. Therefore, building a qualitative relation between the relative gravitational path integral and the PW mechanism opens a way to use existing results in the quantum reference frame as a guide to study the former. The other way around, the gravitational path integral produces certain results which is not naturally expected solely from the quantum reference frame point of view. The problem of dimension is such an example. Therefore, this relation helps us take results from gravitational path integral as inputs for studying quantum reference frame, which opens a new direction also for the latter. More details will be illustrated in section \ref{sec:discussion}. 

The paper is organized as follows. In section \ref{sec:two_issues}, we use a 1D model of gravitational path integral to explain how the problem of time (dimension) arises from summing over metrics (topologies). In section \ref{sec:PW}, we revisit the PW mechanism, see how it resolves the two problems in a purely quantum mechanical setup, and contrast a hierarchy between an observer and a timekeeper. In section \ref{sec:observer_timekeeper}, we present a general formulation of the gravitational path integral relative to an observer and that relative to a timekeeper, 
show the latter furnishes an observer-dependent generalization of holography, and explain their implementation in the 1D gravity model. In section \ref{sec:3D_gravity}, we construct a 3D model, where all the features are incorporated and the relation between the relative gravitational path integral to holographic correspondences becomes transparent. We will also discuss some basic features of the gravitational path integral with multiple observers. In section \ref{sec:discussion}, we summarize the results and discuss future directions based on the relations between the relative gravitational path integral and the quantum reference frame. We hope this paper provides a starting point for readers in each field to learn about results in the other, and to merge the two fields of studies.

\section{Two issues, both from gravitational path integral}\label{sec:two_issues}

In this section, we explicitly study the gravitational path integral for a 1D theory. One major advantage of 1D gravitational path integrals is that a Hilbert space can be constructed explicitly in each step, making the gravitational path integral computation transparent in the Hilbert space language. Through this process, we will see how the gravitational path integral leads to the problem of time via summing over metrics, and to the problem of dimension via summing over topologies. 
Throughout the procedure, we would like to keep aware that the gravitational path integral is {\it a priori} merely a process to compute a number for a given boundary condition. The nontrivial point is always to associate a sensible physical meaning to this number by relating it to some Hilbert space.

Let us first review the general formulation in general dimensions, and then proceed to study a specific 1D model. 
The Lorentzian action of the Einstein gravity coupled to matter fields in $(d+1)$-D is 
\begin{align}\label{eq:L_grav_action}
    I_{\rm grav} = \frac{1}{16\pi G_N} \int d^{d+1}x~\sqrt{-g}~(R-2\Lambda) + I_{\rm GH}+ I_{\rm matter},
\end{align}
where $g_{\mu\nu}$ is the metric, $R_{\mu\nu}$ is the Ricci tensor, $\Lambda$ is the cosmological constant. $I_{\rm GH}$ is the Gibbons-Hawking boundary term and $I_{\rm matter}$ is the action for the matter fields.  

\begin{figure}
    \centering
    \includegraphics[width=0.8\linewidth]{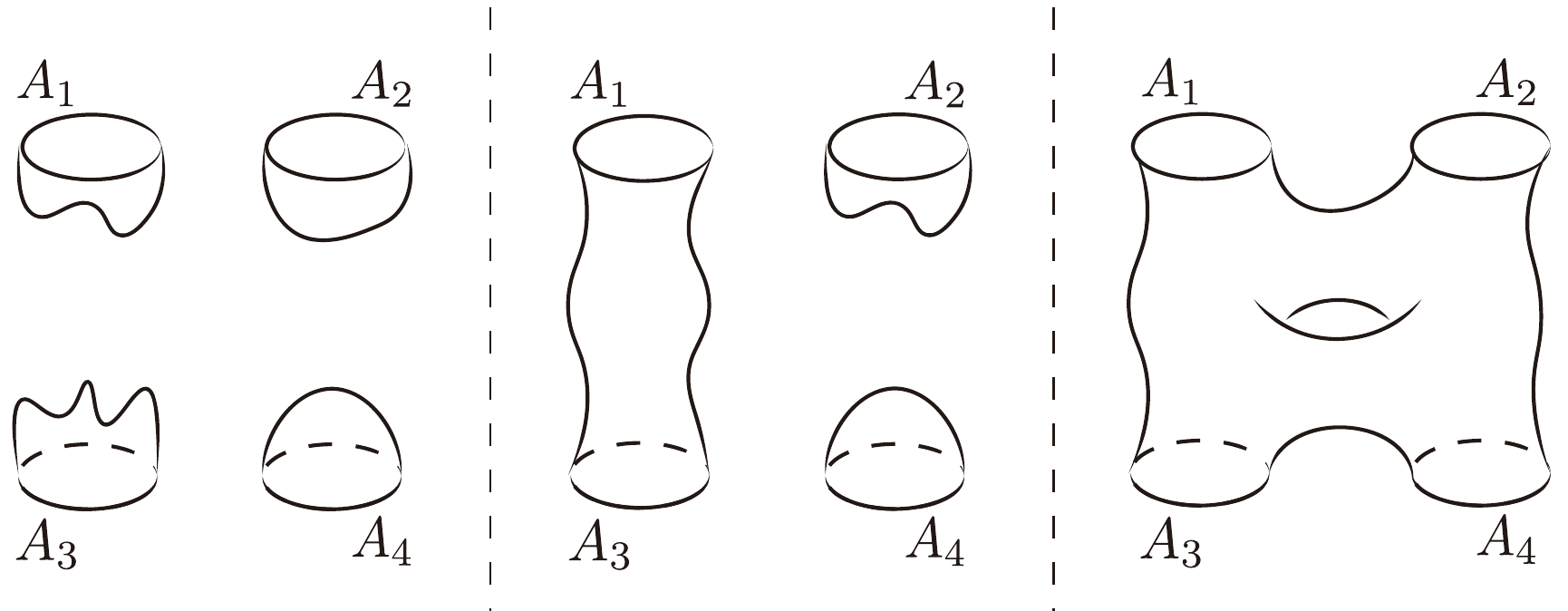}
    \caption{Three examples of geometries with different topologies contributing to the gravitational path integral with four boundaries $G_4$ in the 2D case. }
    \label{fig:2D_manifold}
\end{figure}

In the gravitational path integral in $(d+1)$-D, we specify the boundary conditions on some $d$-D hypersurfaces, and sum over all possible $(d+1)$-D geometries with that boundary condition. 
Such a gravitational path integral computes a set of numbers depending on the boundary condition. Given $N\geq 0$ disconnected hypersurfaces $A_1, A_2, \dots, A_N$, and use $i_{A_1},i_{A_2},\dots,i_{A_N}$ to denote the boundary condition, including the induced metrics and the matter field configurations, imposed on $A_1, A_2, \dots, A_N$. We may denote the set of numbers computed by the gravitational path integral as 
\begin{align}
    G_{N}(i_{A_1},i_{A_2},\dots,i_{A_N}) \equiv \sum_{\rm topology} \int {\mathcal{D}g_{\mu\nu}} \int \mathcal{D}\Phi \exp(iI_{\rm grav}), 
\end{align}
where the sum is taken over all the geometries (including all the topologies and all the metrics $g_{\mu\nu}$) and all the matter field configurations $\Phi$ compatible with the boundary conditions imposed on the $N$ hypersurfaces. Some examples of geometries for $N=4$ appearing in this sum are shown in figure \ref{fig:2D_manifold}. 
Assuming such numbers are computable, though it is not always the case, the central challenge of the gravitational path integral is to associate them to some meaningful physical quantities. 

While most of the results below do not require the explicit form of $I_{\rm matter}$ in \eqref{eq:L_grav_action}, we would like to consider $M$ species of free scalar fields, just to be concrete,
\begin{align}
    I_{\rm matter} &= - \sum_{i=1}^M \frac{1}{2}\int d^{d+1}x~\sqrt{-g}~ \left(\partial_\mu \phi_i \partial^\mu \phi_i + m^2 \phi_i^2 \right) \nonumber \\
    &\equiv -  \frac{1}{2}\int d^{d+1}x~\sqrt{-g}~ \left(|\partial_\mu \vec{\phi}|^2 + m^2 |\vec{\phi}|^2 \right). 
\end{align}
Here, $m$ is the mass for each scalar field, and $\vec{\phi} = (\phi_1, \phi_2, \dots, \phi_M)$.

When $d=0$, i.e. in the case of 1D gravity, we may write the only metric component as 
\begin{align}
    g_{tt}(t) = -e(t)^2. 
\end{align}
The action is reduced to 
\begin{align}\label{eq:1Daction}
    I_{\rm grav} = \frac{1}{2} \int dt \left(\frac{1}{e}\left|\frac{d\vec{\phi}}{dt}\right|^2 - em^2|\vec{\phi}|^2 - \frac{e\Lambda}{4\pi G_N}\right),
\end{align}
where $e(t)$ is often called the einbein field \cite{Polchinski98} in the context of the worldline formulation of quantum field theories (QFT). 

\begin{figure}
    \centering
    \includegraphics[width=0.9\linewidth]{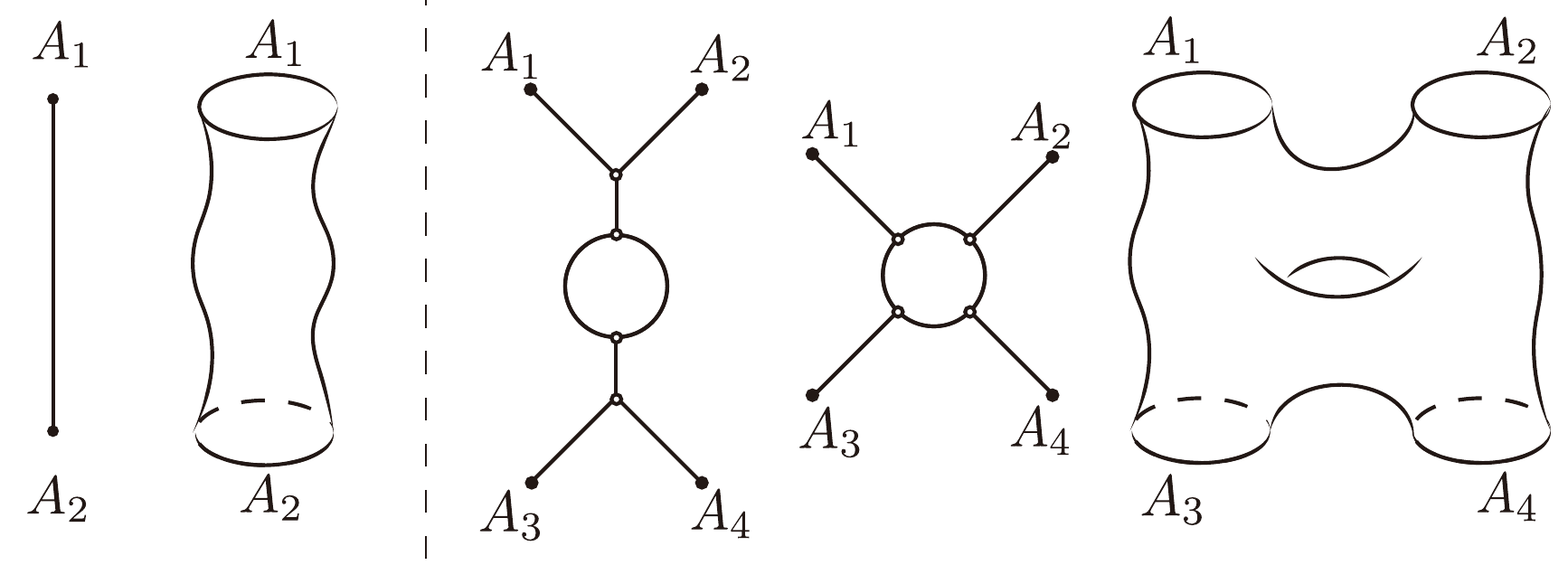}
    \caption{The left shows a 1D manifold connecting two points and its 2D analogue , and the right shows two networks connecting four (external) point and their 2D analogue. In this section, we only consider summing over 1D manifolds.}
    \label{fig:1D_manifold_network}
\end{figure}

In the 1D gravitational path integral, the 0D boundary condition is uniquely specified by the number of disconnected pieces, i.e. points, and the matter field configuration on them. The 1D geometries to be summed over can include manifolds and networks, as shown in figure \ref{fig:1D_manifold_network}. In this section, we consider the most simple case: given a set of points and the field configuration on them, we only sum over 1D manifolds with that boundary condition. This means the basic element is the path integral over a line segment connecting two end points. Let us firstly study them in more details. 

In the following, we will start from summing over matter fields with fixed geometry, then sum over metrics with fixed topology. We will sum over the topology in the end.

\subsection{Kinematic Hilbert space from sum over matter fields} 

Consider the 1D action \eqref{eq:1Daction} on a line segment parameterized by $t'\in[0,1]$. Noticing that \eqref{eq:1Daction} has a gauge redundancy associated to the reparameterization $t' \longmapsto f(t')$, we may perform the gauge fixing to the proper time gauge $e(t') = T$, under which the action can be rewritten as 
\begin{align}\label{eq:1Daction_line}
    I_{\rm grav} &= \frac{1}{2} \int_0^1 dt' \left(\frac{1}{e}\left|\frac{d\vec{\phi}}{dt'}\right|^2 - em^2|\vec{\phi}|^2 - \frac{e\Lambda}{4\pi G_N}\right) \nonumber\\
    &= \int_0^1 dt'  \left(\frac{1}{T}\left|\frac{d\vec{\phi}}{dt'}\right|^2 - Tm^2|\vec{\phi}|^2 - \frac{T\Lambda}{4\pi G_N}\right)    = \int_0^T dt  \left(\left|\frac{d\vec{\phi}}{dt}\right|^2 - m^2|\vec{\phi}|^2 - \frac{\Lambda}{4\pi G_N}\right),
\end{align}
where in the last line we have changed the integral variable. 

The Hamiltonian obtained from this action is 
\begin{align}\label{eq:ADM_Hamiltonian}
    H = \frac{1}{2}|\vec{\pi}|^2 + \frac{1}{2}m^2 |\vec{\phi}|^2 + \frac{\Lambda}{8\pi G_N}, 
\end{align}
where $|\vec{\pi}|$ is the canonical momentum associated to $|\vec{\phi}|$. 
Set the gravity aside for a moment and perform the canonical quantization by imposing $[\phi_i,\pi_j] = i \delta_{ij}$. When the species number $M=1$, the Hamiltonian \eqref{eq:ADM_Hamiltonian} is nothing but that of a harmonic oscillator with an offset, 
\begin{align}
    H = \sum_{n=0}^{\infty} \left(E^{(n)} +  \frac{\Lambda}{8\pi G_N}\right) |n\rangle\langle n|, 
\end{align}
where 
\begin{align}
    E^{(n)} = \left(n+\frac{1}{2}\right)m.
\end{align}
Let us denote the Hilbert space of such a harmonic oscillator $\mathcal{H}_{\rm ho}$.

Accordingly, the Hilbert space for $M$ species is a tensor product of $M$ copies of $\mathcal{H}_{\rm ho}$. Let us call this the kinematic Hilbert space, 
\begin{align}
    \mathcal{H}_{\rm kin} = \mathcal{H}_{\rm ho1} \otimes \mathcal{H}_{\rm ho2} \otimes\cdots \otimes \mathcal{H}_{{\rm ho}M}. 
\end{align}
The Hilbert space of the path-integral quantized gravity will be constructed based on $\mathcal{H}_{\rm kin}$, 
and the reason of this naming will become clear soon. The Hamiltonian can be written as 
\begin{align}
    H = \sum_{n_1=0}^{\infty}\cdots\sum_{n_M=0}^{\infty} \left(E^{(n_1)}+\cdots+E^{(n_M)} +  \frac{\Lambda}{8\pi G_N}\right) |n_1\dots n_M\rangle\langle n_1\dots n_M|, 
\end{align}
in this case.

Let us call the two endpoints of the line segment $A$ and $B$ respectively. Imposing $\vec{\phi}(0) = \vec{\phi}_A$ and $\vec{\phi}(T) = \vec{\phi}_B$ and performing the path integral over the matter field configurations while leaving the geometry unchanged against the action \eqref{eq:1Daction_line}, one gets the transition amplitude on $\mathcal{H}_{\rm kin}$,\footnote{Strictly speaking, $\ket{\vec{\phi}_A}$ and $\ket{\vec{\phi}_B}$ are not normalizable states on $\mathcal{H}_{\rm kin}$. However, we will apply this innocuous abuse of terminologies. }
\begin{align}
    \int_{\vec{\phi}(0) = \vec{\phi}_A}^{\vec{\phi}(T) = \vec{\phi}_B} \mathcal{D}\vec{\phi} \exp\left(\frac{i}{2}\int_0^T dt  \left(\left|\frac{d\vec{\phi}}{dt}\right|^2 - m^2|\vec{\phi}|^2 - \frac{\Lambda}{4\pi G_N}\right) \right) = \braket{\vec{\phi}_B|e^{-iTH}|\vec{\phi}_A},
\end{align}
which is a standard result of the Feynman path integral. 

As a summary, the matter field path integral is {\it a priori} a process to compute a number depending on the boundary condition, and what makes it meaningful is that this number can be identified as an inner product $\braket{\vec{\phi}_B|e^{-iTH}|\vec{\phi}_A}$ on $\mathcal{H}_{\rm kin}$. This $\mathcal{H}_{\rm kin}$ is the first Hilbert space we have encountered so far.

\subsection{Constrained Hilbert space from sum over metrics and the problem of time}
Let us then proceed to sum over all the metrics of line segments connecting two points $A$ and $B$, where the matter configuration is $\vec{\phi}_{A(B)}$ on $A(B)$. Summing over both the matter field configurations and metrics, the path integral is formally computed as  
\begin{align}\label{eq:F_AB}
    G_2(\vec{\phi}_A,\vec{\phi}_B) \propto & \int \mathcal{D}e \mathcal{D}\vec{\phi} \exp\left(\frac{i}{2}\int_0^1 dt  \left(\frac{1}{e}\left|\frac{d\vec{\phi}}{dt}\right|^2 - em^2|\vec{\phi}|^2 - \frac{e\Lambda}{4\pi G_N}\right) \right) \nonumber\\
    \propto&  \int_C dT  \int_{\vec{\phi}(0) = \vec{\phi}_A}^{\vec{\phi}(T) = \vec{\phi}_B} \mathcal{D}\vec{\phi} \exp\left(\frac{i}{2}\int_0^T dt  \left(\left|\frac{d\vec{\phi}}{dt}\right|^2 - m^2|\vec{\phi}|^2 - \frac{\Lambda}{4\pi G_N}\right) \right) \nonumber\\
    = & \int_C dT \braket{\vec{\phi}_B|e^{-iTH}|\vec{\phi}_A}. 
\end{align}
Now, note that the integral contour $C$ is not determined {\it a priori}. There is a degree of freedom to choose it by hand. The most natural choice is to sum over all possible lengths and time orientations, i.e. $-\infty < T < \infty$, with which 
\begin{align}
    G_2(\vec{\phi}_A,\vec{\phi}_B) \propto \int_{-\infty}^{\infty} dT \braket{\vec{\phi}_B|e^{-iTH}|\vec{\phi}_A} \propto \braket{\vec{\phi}_B|\delta(H)|\vec{\phi}_A}.
\end{align}
Interpreting this as a wave function realizing the state 
\begin{align}
    \ket{\Psi} \propto \int \mathcal{D}\vec{\phi}_B ~\ket{\vec{\phi}_B}G_2(\vec{\phi}_A,\vec{\phi}_B).
\end{align}
in $\mathcal{H}_{\rm kin}$, it immediately follows that 
\begin{align}\label{eq:WDW_sec2}
    H\ket{\Psi} = 0,  
\end{align}
which is nothing but the WDW equation causing the problem of time. 

In short, states (in $\mathcal{H}_{\rm kin}$) prepared by a line path integral summing over metrics do not have an explicit time evolution under the Hamiltonian $H$. This result was originally derived by Halliwell in \cite{Halliwell88}, where the main focus was the 1D path integral for minisuperspace model, and later extended to higher dimensions in \cite{HH90}, where summing over the lapse function leads to the Hamiltonian constraint \eqref{eq:WDW_sec2}. We will comment on the higher dimensional case in section \ref{sec:higherD}.

In the following, we list out some important aspects of the path integral studied in this section. Also, we have not explained why $-\infty < T < \infty$ is the best contour we could choose except for saying it is natural. Hence, we will also provide different arguments supporting the contour choice. 

\subsubsection{Constrained Hilbert space and group averaging} 
Under a proper normalization, the path integral realizes the projector $\Pi_{0}$ onto the sub-Hilbert space of $\mathcal{H}_{\rm kin}$ collecting all the states satisfying $H\ket{\Psi} = 0$. 
\begin{align}\label{eq:projector}
    \int \mathcal{D}\vec{\phi}_A \mathcal{D}\vec{\phi}_B~ |\vec{\phi}_A\rangle\langle\vec{\phi}_B| G_2(\vec{\phi}_A,\vec{\phi}_B) = \Pi_{0}. 
\end{align}
The other way around, we can use \eqref{eq:projector} to define the normalization of $G_2$.
We can construct a new inner product using this projector. For $\ket{\Psi_A}, \ket{\Psi_B} \in \mathcal{H}_{\rm kin}$, we define the constrained inner product as 
\begin{align}
    (\Psi_B,\Psi_A)_{\rm con} \equiv \braket{\Psi_B|\Pi_{0}|\Psi_A}. 
\end{align}
This new inner product induces a new Hilbert space which can be regarded as a sub-Hilbert space of $\mathcal{H}_{\rm kin}$ collecting all the vectors satisfying $H\ket{\Psi} = 0$. Let us call this the constrained Hilbert space $\mathcal{H}_{\rm con}$. 

This is nothing but the Dirac quantization for constrained systems, i.e. ``quantize then constrain". In our case, the WDW equation is regarded as the gauge constraint, and the construction of the constrained inner product $(\Psi_B,\Psi_A)_{\rm con}$ involves an integral over the elements in the gauge group generated by $H$, $\int_{-\infty}^{\infty} dT~e^{-i TH}$. This way of constructing a gauge invariant Hilbert space $\mathcal{H}_{\rm con}$ from the kinematic Hilbert space $\mathcal{H}_{\rm kin}$ is called the group averaging, proposed by Marolf \cite{Marolf00}. 

From this perspective, summing over metrics with the contour $-\infty <T < \infty$ constructs a constrained Hilbert space $\mathcal{H}_{\rm con}$ satisfying the WDW constraint. 
This is the second Hilbert space we have encountered so far. 

In the literature of Dirac quantization, the constrained Hilbert $\mathcal{H}_{\rm con}$ here is often called the physical Hilbert space and denoted as $\mathcal{H}_{\rm phys}$. This is because in cases of standard gauge theories, this is the Hilbert space of the physical states. We, however, would like to reserve this terminology. The reason is because we have not considered summing over topologies yet to this point, which is another essential element of the gravitational path integral. 

\subsubsection{No-boundary wave function and no-boundary density matrix}

\begin{figure}
    \centering
    \includegraphics[width=0.95\linewidth]{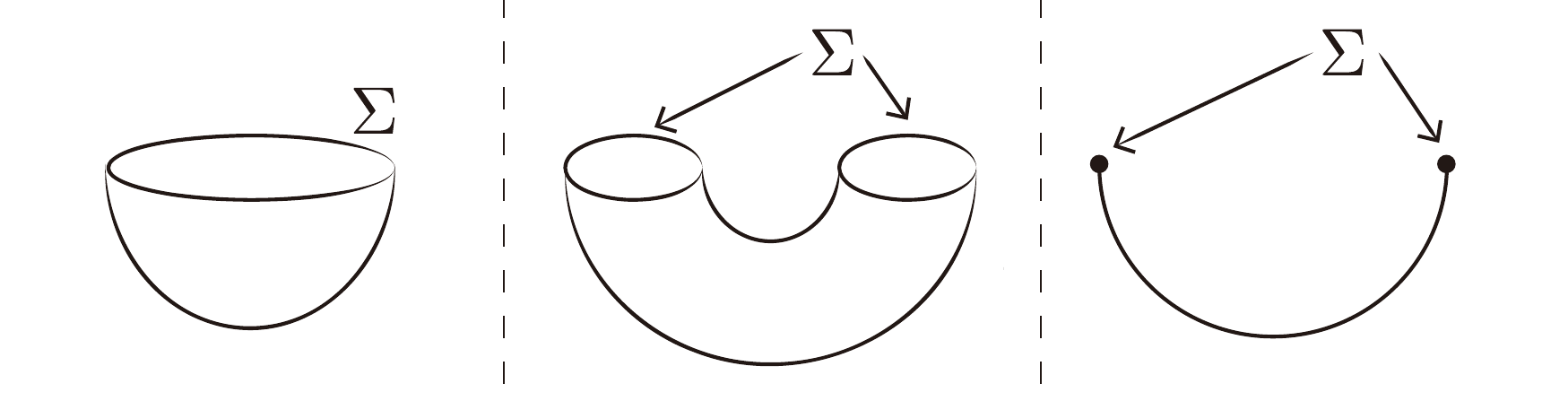}
    \caption{The no-boundary wave function is defined by summing over geometries and field configurations with a hypersurface $\Sigma$ as the boundary, on which the configurations are fixed. Three examples of geometries contributing to some no-boundary wave functions in different dimensions are shown. The left shows a 2D geometry where $\Sigma$ is a circle. The middle shows a 2D geometry where $\Sigma$ is two circles. The right shows a 1D line segment where $\Sigma$ is two points, and this is what is considered in the main text.}
    \label{fig:NB}
\end{figure}

Before moving on to summing over topologies, let us write down some interesting states with the elements obtained so far. 

The first state we would like to consider is the Hartle-Hawking (HH) no-boundary wave function \cite{HH83}, which is defined by fixing data on a spatial hypersurface $\Sigma$ and sum over all the geometries satisfying $\partial \mathcal{M} = \Sigma$ as well as the matter field configurations. In the 1D theory we are considering, $\Sigma$ has to include at least two points $A$ and $B$ to receive a nontrivial contribution.  
Let us consider the case where the spatial slice consists two points $A$ and $B$. See figure \ref{fig:NB} for a sketch. The (unnormalized) no-boundary wave function $\ket{\rm NB}_{AB} \in \mathcal{H}_{{\rm kin},A} \otimes \mathcal{H}_{{\rm kin},B}$ reads 
\begin{align}\label{eq:NB_AB}
    \ket{\rm NB}_{AB} = \int \mathcal{D}\vec{\phi}_A \mathcal{D}\vec{\phi}_B~G_2(\vec{\phi}_A,\vec{\phi}_B) |\vec{\phi}_A\rangle_{{\rm kin}, A}|\vec{\phi}_B\rangle_{{\rm kin}, B} = \sum_i \ket{i}_{{\rm con},A} \otimes \ket{i}_{{\rm con},B}, 
\end{align}
where $\ket{i}_{\rm con}$'s denote an orthonormal basis of $\mathcal{H}_{\rm con}$. In other words, the no-boundary wave function turns out to be a maximally entangled state with respect to $\mathcal{H}_{{\rm con},A} \otimes \mathcal{H}_{{\rm con},B}$. $\ket{\rm NB}_{AB}$ satisfies two sets of WDW equations
\begin{align}\label{eq:two_constraints}
    \left(H_A\otimes I_B\right) \ket{\rm NB}_{AB} = 0, ~~~\left(I_A\otimes H_B\right) \ket{\rm NB}_{AB} = 0, 
\end{align}
acting independently on the two sides of the universe. 

\begin{figure}
    \centering
    \includegraphics[width=0.9\linewidth]{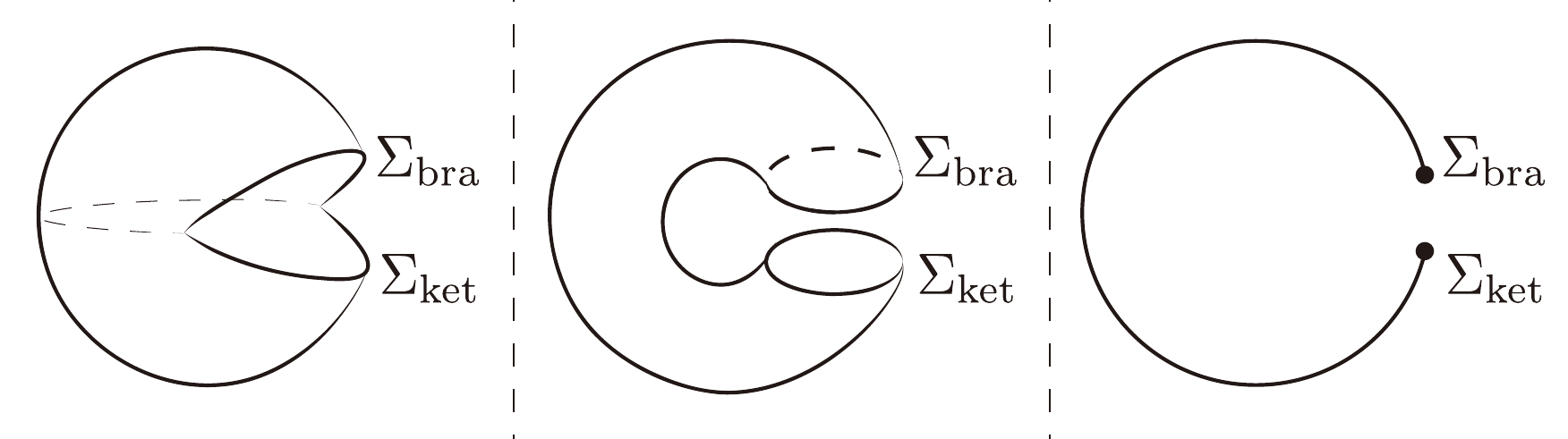}
    \caption{The no-boundary density matrix is defined by summing over geometries and field configurations with two codimension-1 hypersurfaces $\Sigma_{\rm bra}$ and $\Sigma_{\rm ket}$ as the boundary, on which the configurations are fixed and identified as the matrix element. Three examples of geometries contributing to some no-boundary density matrices in different dimensions are shown. The left shows a 2D pacman geometry where both $\Sigma_{\rm bra}$ and $\Sigma_{\rm ket}$ are line segments. The middle shows a 2D bra-ket wormhole geometry \cite{CGM20} where both $\Sigma_{\rm bra}$ and $\Sigma_{\rm ket}$ are circles. The right shows a 1D line segment where both $\Sigma_{\rm bra}$ and $\Sigma_{\rm ket}$ are points, and this is what is considered in the main text.}
    \label{fig:NBDM}
\end{figure}

The second state we would like to consider is the Ivo-Li-Maldacena (ILM) no-boundary density matrix proposed in \cite{ILM24}, which is defined by specifying data on two spatial slices $\Sigma_{\rm bra}$ and $\Sigma_{\rm ket}$, one regarded as bra and the other regarded as ket, and sum over all the geometries $\mathcal{M}$ satisfying $\partial \mathcal{M} = \Sigma_{\rm bra} \cup \Sigma_{\rm ket}$, as well as the matter field configurations. 
See figure \ref{fig:NBDM} for a sketch.
The most simple case is when both $\Sigma_{\rm bra}$ and $\Sigma_{\rm ket}$ are a point. 
The (unnormalized) no-boundary density matrix then reads 
\begin{align}\label{eq:NBden_2p}
    \rho_{\rm NB} = \int \mathcal{D}\vec{\phi}_A \mathcal{D}\vec{\phi}_B~G_2(\vec{\phi}_A,\vec{\phi}_B) |\vec{\phi}_A\rangle\langle\vec{\phi}_B| = \sum_i |i\rangle_{\rm con} \langle i|_{\rm con} = I_{\rm con}, 
\end{align}
which is the identity operator (the maximally mixed state after normalization) on $\mathcal{H}_{\rm con}$. The no-boundary density matrix satisfies the WDW constraint 
\begin{align}
    [H, \rho_{\rm NB}] = 0, 
\end{align}
and hence does not have an explicit time evolution, either. The form \eqref{eq:NBden_2p} is technically identical to a recent result derived in \cite{BKU25}.

There is an extremely nice property of the no-boundary wave function and the no-boundary density matrix computed above.
That is, 
the no-boundary density matrix can be obtained by tracing out one side of the no-boundary state as a subsystem: 
\begin{align}\label{eq:trace_NB}
    \rho_{{\rm NB},A} = \Tr_B~ |{\rm NB}\rangle_{AB} \langle {\rm NB} |_{AB} = I_{{\rm con}, A}. 
\end{align}
Note that this property is not trivial at all in the gravitational path integral as it may look like. 
This is because the computation of $\rho_{{\rm NB},A}$ involves only one $\int dT$ while that of $\Tr_B~ |{\rm NB}\rangle_{AB} \langle {\rm NB} |_{AB}$ involves two. 
Indeed, as we will see later, generic integral contour choices will violate \eqref{eq:trace_NB}, and the analogue of this equality no longer holds once we start summing over topologies in gravitational path integrals with more than two end-points. While the gravitational path integral does not {\it a priori} admit a quantum mechanical interpretation, as emphasized at the beginning of this section, we would like to design it in a way such that standard quantum mechanical rules apply. In this sense, the property \eqref{eq:trace_NB} provides us a good motivation to pick up the contour $-\infty < T < \infty$. 

From another point of view, properties like \eqref{eq:trace_NB} may be used to ``bootstrap" measures in the gravitational path integral, which is not usually given {\it a priori}. It would be interesting to explore this direction in the future.


\subsubsection{Other integral contours}
For completeness, we briefly comment on some other choices of the integral contour for the einbein field.

One choice is to take $0 \leq T < \infty$ to sum over the lengths with a specific time orientation, instead of taking $-\infty < T < \infty$ to sum over both the lengths and the orientations as we have done above. This would result in $G_2 (\vec{\phi}_A,\vec{\phi}_B) \propto  \braket{\vec{\phi}_B|\frac{1}{i(H-i\epsilon)}|\vec{\phi}_A}$. This is the choice one makes when formulating the theory of a worldline particle moving in a target Lorentzian space \cite{CMNP85,Strassler92}. 
In this case, the worldline has a canonical time orientation inherited from that of the target space, and the path integral gives the standard Lorentzian propagator. Another choice is to sum over the lengths from $T = 0$ to $i\infty$, which is unoriented and only the lengths are summed up. This is the choice make in the worldline formulation of a Euclidean QFT and would result in $G_2 (\vec{\phi}_A,\vec{\phi}_B) \propto  \braket{\vec{\phi}_B|\frac{1}{H}|\vec{\phi}_A}$, which is the Euclidean propagator. A detailed comparison between these two contours and $-\infty < T < \infty$ can be found in \cite{CMMR21}. Note that the wave functions prepared by these two contours do not satisfy either the WDW constraint or \eqref{eq:trace_NB}.

On the other hand, one may wonder if $-\infty < T < \infty$ (and the contours homologous to it) is the only contour satisfying the WDW constraint. It was found in \cite{Halliwell88} that another independent contour exists in a minisuperspace model. However, this independent contour is model-specific while $-\infty < T < \infty$ prepares states satisfying the WDW constraint in general models, which makes $-\infty < T < \infty$ preferable. See also \cite{BJ24} for a recent investigation of the integral contour for the lapse function.

\subsection{Fundamental Hilbert space from sum over topologies and the problem of dimension}
So far, we have seen how the matter fields provide a kinematic Hilbert space $\mathcal{H}_{\rm kin}$ as a starting point and how summing over metrics brings it down to a smaller constrained Hilbert space $\mathcal{H}_{\rm con}$ subject to the WDW equation. Generically, $\mathcal{H}_{\rm con}$ is more than one-dimensional and accommodates a nontrivial gauge-invariant algebra. 

We are now ready to sum over topologies, and see how this leads to the conclusion that the Hilbert space is is reduced to one-dimensional. 
In this procedure, we will keep taking the stance that the gravitational path integral is {\it a priori} a machinery to compute numbers given a set of boundary conditions, and we will try to find out an appropriate Hilbert space interpretation to it.

\begin{figure}
    \centering
    \includegraphics[width=0.8\linewidth]{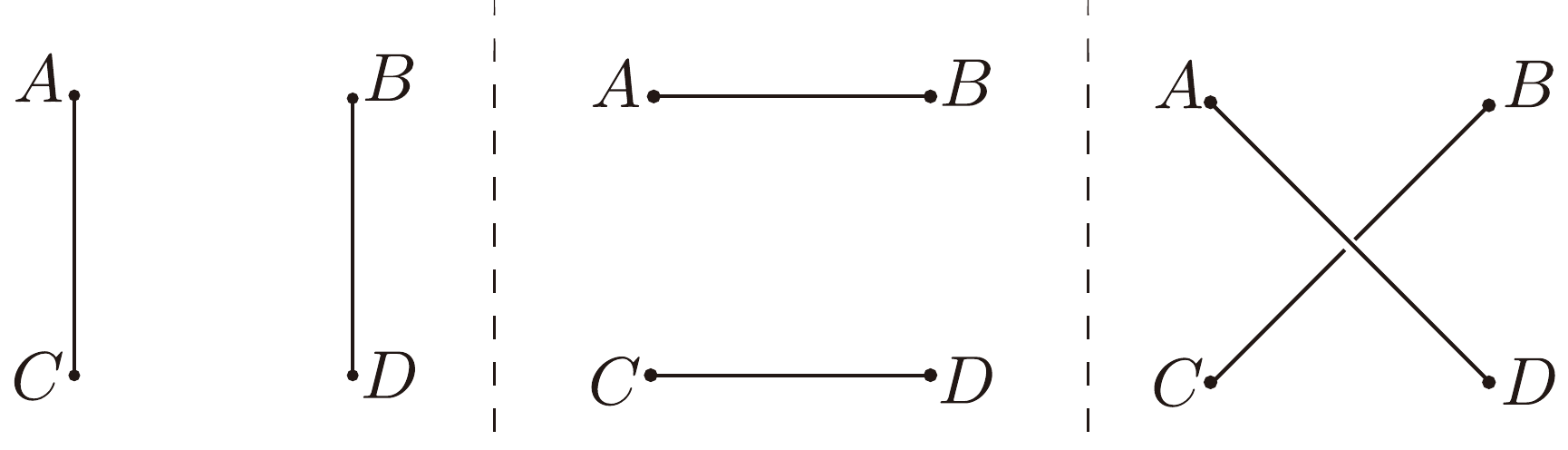}
    \caption{All the three topologies contributing to the 1D gravitational path integral with four disconnected boundaries.}
    \label{fig:four_points}
\end{figure}

The most simple boundary which receives a nontrivial contribution from summing over topologies is four points. Referring figure \ref{fig:four_points}, such a gravitational path integral gives 
\begin{align}
    G_4(\vec{\phi}_A,\vec{\phi}_B, \vec{\phi}_C,\vec{\phi}_D) = G_2(\vec{\phi}_A,\vec{\phi}_B) G_2(\vec{\phi}_C,\vec{\phi}_D) + G_2(\vec{\phi}_A,\vec{\phi}_C) G_2(\vec{\phi}_B,\vec{\phi}_D) + G_2(\vec{\phi}_A,\vec{\phi}_D) G_2(\vec{\phi}_B,\vec{\phi}_D). 
\end{align}
The purpose of the remaining part of this subsection is to study what conclusions one can get when associating some Hilbert space interpretation to $G_4(\vec{\phi}_A,\vec{\phi}_B, \vec{\phi}_C,\vec{\phi}_D)$ and its higher point counterparts, and see how they lead to the problem of dimension under an appropriate interpretation.

\subsubsection{No-boundary states without ensemble average}
Let us start by considering what happens if we interpret $G_4(\vec{\phi}_A,\vec{\phi}_B, \vec{\phi}_C,\vec{\phi}_D)$ as a no-boundary wavefunction \cite{HH83} or a no-boundary density matrix \cite{ILM24}. 

The corresponding (unnormalized) no-boundary wave function is 
\begin{align}
    \ket{\rm NB}_{ABCD} 
    &= \int \mathcal{D}\vec{\phi}_A \mathcal{D}\vec{\phi}_B \mathcal{D}\vec{\phi}_C \mathcal{D}\vec{\phi}_D
    ~G_4(\vec{\phi}_A,\vec{\phi}_B,\vec{\phi}_C,\vec{\phi}_D) |\vec{\phi}_A\rangle_{{\rm kin}, A}|\vec{\phi}_B\rangle_{{\rm kin}, B} |\vec{\phi}_C\rangle_{{\rm kin}, C}|\vec{\phi}_D\rangle_{{\rm kin}, D} \nonumber\\
    &= \sum_{i,j} \left(|iijj\rangle_{{\rm con}, ABCD} + |ijij\rangle_{{\rm con}, ABCD} + |ijji\rangle_{{\rm con}, ABCD} \right), 
\end{align} 
and the corresponding (unnormalized) no-boundary density matrix is 
\begin{align}
    \rho_{{\rm NB},AB} = \sum_{i,j} \left(|ii\rangle_{{\rm con}, AB}\langle jj|_{{\rm con}, AB} + |ij\rangle_{{\rm con}, AB} \langle ij|_{{\rm con}, AB}+ |ij\rangle_{{\rm con}, AB} \langle ji|_{{\rm con}, AB} \right). 
\end{align}
If ${\rm dim} \mathcal{H}_{\rm con} \geq 2$, then we immediately have 
\begin{align}
    &\rho_{{\rm NB},AB} \not\propto |{\rm NB}\rangle_{AB} \langle{\rm NB}|_{AB}, \nonumber \\
    &\rho_{{\rm NB},AB} \not\propto \Tr_{CD}\left(|{\rm NB}\rangle_{ABCD} \langle{\rm NB}|_{ABCD}\right) \not\propto |{\rm NB}\rangle_{AB} \langle{\rm NB}|_{AB}. 
\end{align}
The first equation means that, for an observer who can access both $A$ and $B$, the rule of computing the state of the universe via the no-boundary wave function \cite{HH83} and that via the no-boundary density matrix \cite{ILM24} do not give the same result. 
The second equation implies that, the state of the universe seen by an observer having access to $A$ and $B$ is not identical to that seen by an observer having access to $A,B,C$ and $D$ and then restricting on $A,B$. This is a primitive sign that the gravitational path integral is observer-dependent, although we have not given an rigorous formulation of an ``observer" yet.

It is interesting to note that, if we persist this interpretation, then one finds that 
\begin{align}
    \rho_{{\rm NB}A_1A_2...A_m}  &\propto |{\rm NB}\rangle_{A_1A_2...A_m} \langle{\rm NB}|_{A_1A_2...A_m} \nonumber \\
    &\propto \Tr_{A_{m+1}...A_n}\left(|{\rm NB}\rangle_{A_1A_2...A_m...A_n} \langle{\rm NB}|_{A_1A_2...A_m...A_n}\right)
\end{align}
hold if and only if 
\begin{align}
    {\rm dim} \mathcal{H}_{\rm con} = 1.
\end{align}
However, even in this case, it is not clear how the normalization of the states can be properly performed in terms of the gravitational path integral.

\subsubsection{No-boundary states with ensemble average}\label{sec:NBwithEnsemble}
Let us then proceed to consider an alternative interpretation. It has been argued in \cite{Coleman88,GS88,MM20} that the gravitational path integral does not have enough resolution to serve as a UV complete theory, but should be instead interpreted as computing averaged quantities over an (possibly unknown) ensemble. 
Indeed, there exist concrete models where an exact gravitational path integral is dual to an ensemble of quantum systems \cite{SSS19}. 

Under this interpretation, the argument below is essentially a rephrasing of that given in \cite{MM20,UWZ24,UZ24}, with the interpretation in terms of the no-boundary matrices a slight new ingredient. 

As a simple example under this interpretation, when considering the no-boundary density matrix, instead of regarding $G_2(i,j)$ as $(\rho_{\rm NB})_{ij}$ as what we did in \eqref{eq:NBden_2p}, we are supposed to regard it as 
\begin{align}
    \overline{(\rho_{\rm NB})_{ij}} = G_2(i,j)
\end{align}
instead, where the overline means that an ensemble average is taken over an undetermined set of microscopic theories. 

Similarly, $G_4(i,j,k,l)$ can be related to several different objects. Regarding the four end points as two copies of $\Sigma_{\rm bra}$ and $\Sigma_{\rm ket}$, we have
\begin{align}
    \overline{(\rho_{\rm NB})_{ij}(\rho_{\rm NB})_{kl}} = G_4(i,j,k,l). \label{eq:averaged_squared}
\end{align}
Regarding two end points $A,B$ as $\Sigma_{\rm bra}$, and the other two as $\Sigma_{\rm ket}$, we have 
\begin{align}\label{eq:G4_2}
      \overline{(\rho_{{\rm NB},AB})_{ijkl}} = G_4(i,j,k,l).
\end{align}
Regarding two end points $A,B$ as $\Sigma$, and the other two as a copy of them, we have
\begin{align}\label{eq:G4_3}
    \overline{((|{\rm NB}\rangle_{AB})_{ij}(\langle {\rm NB}|_{AB})_{kl})} = G_4(i,j,k,l),
\end{align}
where we have introduced the notation $|{\rm NB}\rangle_{AB} = \sum_{i,j}(|{\rm NB}\rangle_{AB})_{ij}\ket{i}_A\ket{j}_B$. The consistency between defining no-boundary states via the no-boundary wave function \cite{HH83}, and that via the no-boundary density matrix \cite{ILM24} is thus recovered. 

Let us look at \eqref{eq:averaged_squared} more carefully. It leads to 
\begin{align}
    \overline{\left(\Tr\rho_{\rm NB}\right)^2} = \sum_{i,j} \overline{(\rho_{\rm NB})_{ii}(\rho_{\rm NB})_{jj}} = \sum_{i,j} \overline{(\rho_{\rm NB})_{ij}(\rho_{\rm NB})_{ji}} = \overline{\Tr\left(\rho_{\rm NB}^2\right)}. 
\end{align} 
where we have used the permutation symmetry of $G_4(i,j,k,l)$. Similarly, the permutation symmetry of $G_{2n}$ leads to 
\begin{align}
    \overline{\left(\Tr\rho_{\rm NB}\right)^n} = \overline{\Tr\left(\rho_{\rm NB}^n\right)}. 
\end{align} 
and the permutation symmetry of $G_{4n}$ leads to 
\begin{align}
    \overline{\left(\Tr(\rho_{\rm NB}^n) - (\Tr\rho_{\rm NB})^n\right)^2} = 0. 
\end{align}
As a result, for each microscopic theory (let us label it by $\alpha$) in the ensemble averaging, 
\begin{align}
    (\Tr \rho_{\rm NB}^{(\alpha)})^n = \Tr (\rho_{\rm NB}^{(\alpha)})^n
\end{align}
Therefore, the no-boundary density matrix defined on a connected spatial slice (in this case, a point) must be rank 1 and hence in a pure state. 

Following a similar discussion, the permutation symmetry of $G_{n}$'s leads to 
\begin{align}
    &\rho^{(\alpha)}_{{\rm NB}} = |{\rm NB^{(\alpha)}}\rangle \langle {\rm NB}^{(\alpha)}|, \\
    &\rho_{{\rm NB},A_1...A_n}^{(\alpha)} = \rho_{{\rm NB},A_1}^{(\alpha)} \otimes ... \otimes \rho_{{\rm NB},A_n}^{(\alpha)}, 
\end{align}
for each member of the ensemble labeled by $\alpha$. This implies the consistency between the no-boundary wave function and the no-boundary density matrix, and the factorization between disconnected universes. 

Moreover, since $G_2(i,j)$ also serves as the averaged $ij$ element of the Gram matrix with respect to the inner product $(i,j)_{\rm con}$ under this interpretation, its being rank one implies that, for each microscopic theory labeled by $\alpha$ in the ensemble, only one state survives under the gravitational inner product modified by the contributions from different topologies. This new inner product leads to a more fundamental Hilbert space, which is one-dimensional in each member of the ensemble. 
\begin{align}
    \dim \mathcal{H}^{(\alpha)}_{\rm fund} = 1. 
\end{align}
This is the third Hilbert space we have encountered, and it is the space on which the problem of dimension arises. 

\subsection{Summary and notes on higher dimensions}\label{sec:higherD}
So far, we have explained how the problem of time and the problem of dimension arise in a 1D gravitational path integral. 

We started by constructing the kinematic Hilbert space $\mathcal{H}_{\rm kin}$ arising from summing over the matter field, then we saw how summing over metrics when restricting to the interval topology imposes the Hamiltonian constraint on $\mathcal{H}_{\rm kin}$, resulting in a smaller constrained Hilbert space $\mathcal{H}_{\rm con}$. The problem of time arises in this procedure. 
Then we take the ensemble average interpretation into account. The permutation symmetry of the gravitational path integral with multi-boundaries $G_n$ further restricts $\mathcal{H}_{\rm kin}$ to the one-dimensional fundamental Hilbert space $\mathcal{H}_{\rm fund}$ for each member of the average ensemble. The problem of dimension arises in this step. 

\subsubsection*{The origin of the two problems and how to avoid them}

Mathematically, the one-dimensional nature of the fundamental Hilbert space $\mathcal{H}_{\rm fund}$ and the factorization of the no-boundary density matrix between the replicas of a connected universe were derived from the permutation symmetry of $G_n$. In the gravitational path integral we considered, this permutation symmetry originates from summing over all the topologies: all the end points are not special from each other, and hence all the topologies are summed over in a balanced way. 

It is immediately clear that the problem of dimension occurs because different topologies are summed over {\it too nicely}. 
Therefore, the problem of dimension is not robust against deformation of the sum. 
Indeed, if the topologies were summed over in slightly unbalanced way or only certain topologies were picked up in the sum, such that the permutation invariance of $G_n$ were broken, then the fundamental Hilbert space would have a non-trivial dimension. 
The question is then to find a physically sensible reason to sum over the topologies in an unbalanced way. 

This is also the reason why there exist several different proposals \cite{BNU23,HUZ25,AAIL25} to deal with the problem dimension. They all break the permutation invariance of $G_n$, but in different ways. 

Another remark to make is that, although we first summed over metrics and then the topologies to obtain $\rm dim~\mathcal{H}_{\rm fund} =1$ above, summing over topologies and orientations (without summing over the metrics) is in fact sufficient.\footnote{In the literature, the Hilbert space starting from which the wormhole corrections are to include is often called the effective Hilbert space $\mathcal{H}_{\rm eff}$ (See, e.g. \cite{AEHPV22}). In our case, we have taken $\mathcal{\rm con}$ to play the role of $\mathcal{H}_{\rm eff}$, but we could have taken e.g. $\mathcal{H}_{\rm kin}$ as $\mathcal{H}_{\rm eff}$, from which we will again reach to a problem of dimension after summing over topologies. This aspect also contrasts the origin of the problem of time and that of the problem of dimension. } Therefore, the problem of dimension and the problem of time are mathematically independent from each other. 

Similarly, the problem of time occurs because different metrics are summed over {\it too nicely}. Therefore, it is not robust against deformation of the integral. Indeed, if the metrics were summed over in a slightly unbalanced way or only certain metrics were picked up in the sum, then the WDW equation would be broken. 

The question is then to find a physically sensible reason to sum over the metrics in an unbalanced way.


\subsubsection*{Straightforward extension to higher dimensions}

Although we have focused on the 1D gravitational path integral above, all the discussions can be straightforwardly extended to higher dimensions. 

As we have seen above, the problem of dimensions arises from the permutation invariance of the gravitational path integral with multiple boundaries $G_n$. The exactly same argument holds in higher dimensions. 

The case is slightly more complicated for the problem of time. In the 1D case, the WDW equation arises from summing over the only metric component, which reduces to the length (and the orientation) of the line segment. In higher dimensional case, this corresponds to summing over the lapse function appearing in the ADM decomposition. The ADM decomposition of a $(d+1)$-D metric can be written as 
\begin{align}
    ds^2 = -N^2 d\tau^2 + h_{ab} \left(N^a d\tau  + dy^a \right)\left(N^b d\tau  + dy^b \right),
\end{align}
where $h_{ab}(\tau,y^a)$ is the induced metric on the $d$-D slice $\Sigma_{\tau}$ and $N(\tau,y^a)$ is the lapse function. The 1D case is a special case where only the first term in the ADM decomposition is present, and the WDW equation arises from the integral over the lapse function $N(\tau)$ from $-\infty$ to $\infty$. In higher dimensional cases, the integral over the lapse function is a part of summing over metrics. One will again finds that this integral impose the Hamiltonian constraint (WDW equation) on each spatial point $y^a$, and hence leads to the problem of time, although the procedure of constructing $\mathcal{H}_{\rm kin}$ from the ADM Hamiltonian is often challenging. 
See \cite{HH90} for details and \cite{BJ24} for a recent discussion. 

Again, in the higher dimensional cases, the problem of time (dimension) arises from summing over the metrics (topologies) {\it too nicely}. A generic deformation of the path integral or restriction on the integrated configuration space would give a nontrivial time evolution (Hilbert space). The question is to find a physically sensible way for doing so. 

Before moving on to doing so in the gravitational path integral, let us revisit the Page-Wootters mechanism \cite{PW83}, which resolves the two problems in a purely (non-gravitational) quantum mechanical setup. We will then propose an implementation of its analog in the gravitational path integral, making their relations clear in each procedure.

\section{Page-Wootters mechanism revisited}\label{sec:PW}

In this section, we revisit the PW mechanism \cite{PW83,Wootters84} and see how they proposed to solve the two issues, although the problem of dimension was not aware back then. 

The formulation we performed here looks a little bit different from the original work, but they are mathematically equivalent. In particular, some notions which were not distinguished in the original work will be clearly distinguished here. This may look redundant in the purely quantum mechanical setup, but we will see that it is crucial when discussing the gravitational path integral. 
Accordingly, different terminologies from the original work will be used,
but we will comment on the distinctions properly.

\subsection{General formulation}
Let us start from a given kinematic Hilbert space $\mathcal{H}_{\rm kin}$ and the WDW equation under a given Hamiltonian $H$, 
\begin{align}\label{eq:WDW}
    H \ket{\Psi}_{\rm uni} = 0~, 
\end{align}
which indicates the state of the universe does not have an explicit time evolution. We further assume that the solution space of $\eqref{eq:WDW}$ is 1D\footnote{This assumption is not made in the usual discussion of the PW mechanism, but we will see that this assumption is not essential.}, giving arise to both the problem of time and the problem of dimension. 

The spirit of the PW mechanism is to realize that the standard quantum mechanics assumes an observer exists outside the observed system, while in the case of the universe, the observer is a part of the universe, and they are only able to observe the time evolution relative to themselves. This leads to the assumption that the whole kinematic Hilbert space of the universe can be divided into a composition between an observer and their complement, called the environment,\footnote{The observer degrees of freedom here is called the clock in \cite{PW83,Wootters84}, and the environment degrees of freedom here is often simply called the interested system in the literature 
(See, e.g. \cite{HSL19}).} 
\begin{align}
    \mathcal{H}_{\rm kin} = \mathcal{H}_{\rm obs} \otimes \mathcal{H}_{\rm env}~,
\end{align}
and the Hamiltonian $H$ appearing in \eqref{eq:WDW} can be written as 
\begin{align}
    H = H_{\rm obs}\otimes I_{\rm env} + I_{\rm obs} \otimes H_{\rm env} + H_{\rm int} ~.
\end{align} 
These are the major assumptions in the PW mechanism. 

The resulting wave function of the universe is generically an entangled state between the observer and the environment. 
\begin{align}
    \ket{\Psi}_{\rm uni} = \mathcal{N}\sum \lambda_i \ket{\alpha_i}_{\rm obs} \otimes \ket{\beta_i}_{\rm env}, 
\end{align}
where $\mathcal{N}$ is the normalization constant. 

Given a state of the observer $\ket{\psi}_{\rm obs}$, we can define the following observer-to-environment map: 
\begin{align}
    \begin{aligned}
        f_{\rm PW}: \mathcal{H}_{\rm obs} & \longrightarrow \mathcal{H}_{\rm env} \\
        \ket{\psi}_{\rm obs} & \longmapsto f_{\rm PW}\left(\ket{\psi}_{\rm obs}\right) = \frac{\bra{\psi}_{\rm obs} \ket{\Psi}_{\rm uni}}{|\bra{\psi}_{\rm obs} \ket{\Psi}_{\rm uni}|}.
    \end{aligned}
\end{align}
The environment wave function $f_{\rm PW}\left(\ket{\psi}_{\rm obs}\right)$ obtained in this procedure is interpreted as the environment state seen by the observer when the observer themselves are in the state $\ket{\psi}_{\rm obs}$. Although we have only focused on wave functions (pure states), the observer-to-environment map can be defined for mixed states in a similar way, i.e. given a density matrix $\rho_{\rm obs}$, we can define the corresponding (unnormalized) environment state as $\bra{\Psi}_{\rm uni}\rho_{\rm obs} \ket{\Psi}_{\rm uni}$.

Under this interpretation, the effective Hilbert space visible to an observer is nontrivial even though the WDW equation only has one solution $\ket{\Psi}_{\rm uni}$. In particular, it is roughly evaluated by ${\rm rank} \left(\Tr_{\rm env} (|\Psi\rangle_{\rm uni}\langle \Psi|_{\rm uni})\right)$, which is upper bounded by 
\begin{align}
    {\rm rank} \left(\Tr_{\rm env} (|\Psi\rangle_{\rm uni}\langle \Psi|_{\rm uni})\right) \leq \dim \mathcal{H}_{\rm obs}. 
\end{align}
The problem of dimension is hence avoided in this way in the PW framework. As an example, the effective dimension of the visible Hilbert space for a choice of the observer in the 1D gravity studied in section \ref{sec:two_issues}, and its dependence on the Newton constant $G_N$ in the semiclassical limit, are presented in appendix \ref{app:observerEE_1D}.

\begin{figure}
    \centering
    \includegraphics[width=0.9\linewidth]{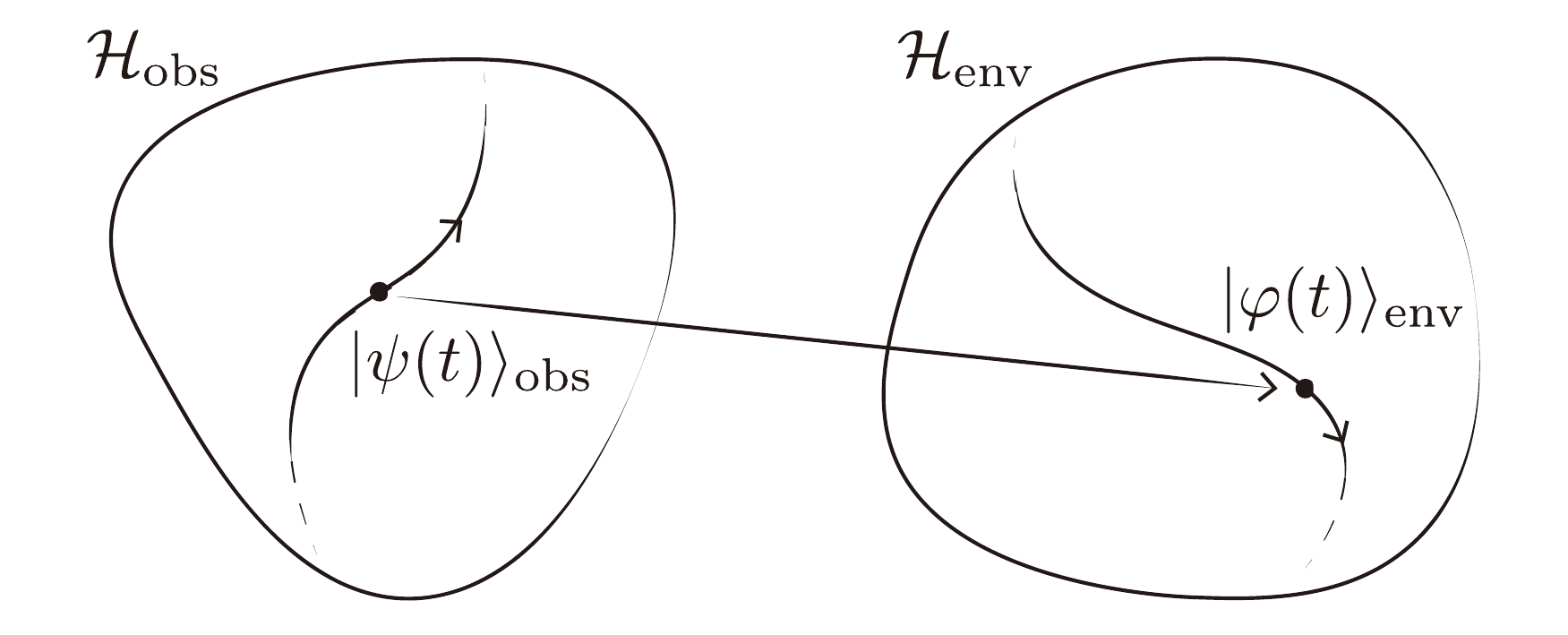}
    \caption{An image of the PW mechanism. The wave function of the universe $\ket{\Psi}_{\rm uni}$ serves as an observer-to-environment map from the observer Hilbert space $\mathcal{H}_{\rm obs}$ to the environment Hilbert space $\mathcal{H}_{\rm env}$. Specifying a history of the observer $\{\ket{\psi(t)}_{\rm obs}\}$ and uplift them to a timekeeper, the PW's map gives a history of the environment. The history of the universe from the timekeeper's perspective is the tensor product between the timekeeper's history and the corresponding environment's history.}
    \label{fig:PWimage}
\end{figure}

Let us then consider a sequence of observer states $\{\ket{\psi(t)}_{\rm obs}\}$ parameterized by $t$, where $t$ runs in a continues or discrete set of real numbers. Performing the observer-to-environment map defined above, we then obtain a corresponding sequence of environment states 
\begin{align}
    \ket{\varphi(t)}_{\rm env} \equiv 
    f(\ket{\psi(t)}_{\rm obs})
    =
    \frac{\bra{\psi(t)}_{\rm obs} \ket{\Psi}_{\rm uni}}
    {|\bra{\psi(t)}_{\rm obs} \ket{\Psi}_{\rm uni}|}. 
\end{align}
The sequence $\{\ket{\psi(t)}_{\rm obs}\}$ is interpreted as a reference of time, i.e. when the observer is in state $\ket{\psi(t')}_{\rm obs}$, we define the time to be $t=t'$. A time evolution of the environment is then retrieved from the observer's perception. We call such a sequence $\{\ket{\psi(t)}_{\rm obs}\}$ a {\it history} of the observer. We will also call an observer with a specified history a {\it timekeeper}, since they can use their own history to refer time. See figure \ref{fig:PWimage} for an image of the PW mechanism (Note that it is just an image which is not precise).

To this point, the choice of the timekeeper's history and the way how the environment time evolution is retrieved look rather artificial and distinct from the standard rule of time evolution in conventional quantum mechanics. To resolve this tension, let us consider the simple case where the interacting term is weak enough to be negligible $\|H_{\rm int}\| \rightarrow0$,\footnote{In gravity, this implies that the observer and the environment are interacting only via gravity (this is where the WDW equation comes from) but not any other forces.} and pick up the history of the observer to be a sequence of states connected by the unitary evolution generated by $H_{\rm obs}$, 
\begin{align}
    \ket{\psi(t)}_{\rm obs} \equiv e^{-itH_{\rm obs}} \ket{\psi(0)}_{\rm obs}~.
\end{align}
The state of the environment at time $t$ from this timekeeper's point of view is  
\begin{align}
    \ket{\varphi(t)}_{\rm env} = \frac{\bra{\psi(t)}_{\rm obs} \ket{\Psi}_{\rm uni}}{|\bra{\psi(t)}_{\rm obs} \ket{\Psi}_{\rm uni}|}~, 
\end{align}
which also has a standard unitary time evolution in quantum mechanics as a consequence of the WDW equation \eqref{eq:WDW}: 
\begin{align}
    \ket{\varphi(t)}_{\rm env} = \frac{\bra{\psi(0)}_{\rm obs} e^{itH_{\rm obs}} \ket{\Psi}_{\rm uni}}{|\bra{\psi(0)}_{\rm obs} e^{itH_{\rm obs}} \ket{\Psi}_{\rm uni}|} = e^{-itH_{env}}\frac{\bra{\psi(0)}_{\rm obs}  \ket{\Psi}_{\rm uni}}{|\bra{\psi(0)}_{\rm obs} \ket{\Psi}_{\rm uni}|} = e^{-itH_{\rm env}} \ket{\varphi(0)}_{\rm env}~. 
\end{align}
As clear from the construction, the time evolution of the environment depends on the choice of the observer and their history. 

Also note that while $\ket{\Psi}_{\rm uni}$ is gauge invariant under \eqref{eq:WDW} and (generically) entangled, the state of the universe at each $t$, $\ket{\psi(t)}_{\rm obs}\otimes\ket{\varphi(t)}_{\rm env}$ is a gauge-dependent notion and a product state. This gauge-dependence of the time evolution is a crucial feature of the PW mechanism. 

For concreteness, let us consider a 2-qubit toy example.


\subsection{A 2-qubit example}
As a simple but concrete example, let both the observer and the environment be a single qubit. Let the Hamiltonian be 
\begin{align}
    H = Z_{\rm obs} \otimes I_{\rm env} + I_{\rm obs} \otimes Z_{\rm env} + \varepsilon \left(X_{\rm obs} \otimes X_{\rm env} - I_{\rm obs} \otimes I_{\rm env}\right)~,  
\end{align}
where $Z = |0\rangle \langle 0| - |1\rangle \langle 1|$ and $X = |0\rangle \langle 1| + |1\rangle \langle 0|$ are Pauli matrices. 

In this case, there is only one state satisfying the WDW equation \eqref{eq:WDW}: 
\begin{align}
    \ket{\Psi}_{\rm uni} = \frac{1}{\sqrt{2}} \left(\ket{0}_{\rm obs}\ket{1}_{\rm env} + \ket{1}_{\rm obs}\ket{0}_{\rm env} \right). 
\end{align}
Therefore, besides the problem of time, there is also the problem of dimension in this setup, although it is an artificial one. 

Consider the weak coupling limit $\varepsilon \rightarrow 0$. 
Let us choose the timekeeper with the following history parameterized by $t$, e.g. 
\begin{align}
    \ket{\psi(t)}_{\rm obs} \equiv e^{-itH_{\rm obs}} \frac{1}{\sqrt{2}} \left( \ket{0}_{\rm obs} + \ket{1}_{\rm obs}\right). 
\end{align}
The PW mechanism inserts that this timekeeper will see the following time evolution of the environment, 
\begin{align}
     \ket{\varphi(t)}_{\rm env} = \frac{\bra{\psi(0)}_{\rm obs} e^{itH_{\rm obs}} \ket{\Psi}_{\rm uni}}{|\bra{\psi(0)}_{\rm obs} e^{itH_{\rm obs}} \ket{\Psi}_{\rm uni}|} = e^{-itH_{env}}\frac{\bra{\psi(0)}_{\rm obs}  \ket{\Psi}_{\rm uni}}{|\bra{\psi(0)}_{\rm obs} \ket{\Psi}_{\rm uni}|} = e^{-itH_{\rm env}} \frac{1}{\sqrt{2}} \left( \ket{1}_{\rm env} + \ket{0}_{\rm env}\right), 
\end{align}
where we have used the WDW equation \eqref{eq:WDW}. In this scenario, although the whole universe stays in a single state, an timekeeper can see a two dimensional Hilbert space, which arises from the entanglement between the observer and the environment in $\ket{\Psi}$. 

We can also consider an alternative timekeeper
\begin{align}
    &\ket{\psi'(t)}_{\rm obs} \equiv e^{-itH_{\rm obs}} \ket{0}_{\rm obs} = e^{-it}\ket{0}_{\rm obs}, \nonumber\\
    & \ket{\varphi'(t)}_{\rm env} = \frac{\braket{\psi'(t)|_{\rm obs}|\Psi}_{\rm uni}}{|\braket{\psi'(t)|_{\rm obs}|\Psi}_{\rm uni}|} = e^{+it}\ket{1}_{\rm env},
\end{align}
where the timekeeper can only see a one-dimensional Hilbert space. 

The visible dimension of the Hilbert space to an observer depends not only how the observer subsystem is chosen, but also on the timekeeper. In both examples above, the dimension of the Hilbert space visible to the timekeeper is no larger than $\dim \mathcal{H}_{\rm obs} =2$. An example where $\dim \mathcal{H}_{\rm obs}$ is infinity is studied in appendix \ref{app:observerEE_1D}. 

\subsection{Summary and remarks}

The setup of the PW mechanism starts from a single static state of the universe $\ket{\Psi}_{\rm uni}$, which entangles all the elements in it. One then identifies a proper subsystem as an {\it observer}, and uses $\ket{\Psi}_{\rm uni}$ to define an {\it observer-to-environment map}, whose image is interpreted as the state of the environment seen by the observer when they are in the given state. Such an observer can see a nontrivial Hilbert space thanks to the entanglement between the observer and the environment. To retrieve a nontrivial time evolution, one specifies a sequence of observer's states called a {\it history} to uplift the observer to a {\it timekeeper}. Such a timekeeper can use their own history to refer time, and hence see a nontrivial time evolution of the environment. In the limit where the explicit interaction $H_{\rm int}$ between the observer and the environment is absent, the standard unitary time evolution of the environment is retrieved.\footnote{This is, however, still different from the standard quantum mechanics. For example, the timekeeper's states corresponding to different time can have nontrivial overlaps $\braket{\psi(t)|_{\rm obs}|\psi(t')}_{\rm obs} \neq 0$ for $t\neq t'$, and the time measured by the timekeeper can be cyclic, which are not what we would expect in the conventional quantum mechanics. A choice of timekeeper fully retrieving the conventional quantum mechanics is called an ideal clock in literature. An ideal clock should satisfy some additional properties, including being infinite dimensional. See for example \cite{SW58,Peres80,UW89}.}

Given a timekeeper with history $\ket{\psi(t)}_{\rm obs}$, the history of the universe is described by $\ket{\psi(t)}_{\rm obs} \otimes \ket{\varphi(t)}_{\rm env}$, which does not satisfy \eqref{eq:WDW} and hence not gauge invariant. This is a crucial feature of the PW mechanism. 

In one word, all the possible  observers, histories and timekeepers are already encoded in the single static state $\ket{\Psi}_{\rm uni}$ in the PW mechanism, and what one needs to do is to pick up a proper one of them. Compared to the original work, we have contrasted the hierarchy between an observer and a timekeeper, which will become important when incorporating their analogs in the gravitational path integral.

\section{Observer/timekeeper in gravitational path integral }\label{sec:observer_timekeeper}

In section \ref{sec:two_issues}, we have seen that the problem of time (dimension) 
arises from summing over metrics (topologies) {\it too nicely} in the gravitational path integral. Let us schematically write this sum as 
\begin{align}\label{eq:fullPI}
    \sum_{\rm topology} \int {\mathcal{D}g} \int\mathcal{D}\Phi \exp(iI_{\rm grav}), 
\end{align}
A generic unbalanced weight in the sum of topologies 
\begin{align}
    \sum_{\rm topology} W({\rm topology}) \int {\mathcal{D}g} \int \mathcal{D}\Phi \exp(iI_{\rm grav}), 
\end{align}
would give a nontrivial Hilbert space, and a generic unbalanced weight in the sum of metrics 
\begin{align}
    \sum_{\rm topology}  \int {\mathcal{D}g} ~W(g)\int \mathcal{D}\Phi \exp(iI_{\rm grav}), 
\end{align}
would lead to a violation of the WDW equation. 

Given the vulnerability of the two problems, the question is more about finding a physically sensible unbalanced weight. Indeed, there exist several different proposals on summing over topologies with unbalanced weights $W({\rm topology})$ to retrieve a nontrivial Hilbert space \cite{BNU23,HUZ25,AAIL25}. On the other hand, proposals on summing over metrics with an unbalanced weigh have not been discussed in the literature. In this section, we will present general proposals for both. 

The proposal is inspired by the PW mechanism revisited in section \ref{sec:PW}, which deals with the two problems in a purely quantum mechanical setup. There, one firstly specifies a subsystem of the universe as the observer, and then specifies a history of them to uplift them to a timekeeper. A corresponding history of the environment, which uses the timekeeper's history as a reference of time, is given by an observer-to-environment map. 

We will perform an analogous construction in the gravitational path integral. We first give rules on how an observer is specified and how it introduces an unbalanced weight in summing over topologies. We then explain how to specify a history of the observer to uplift them to a timekeeper, how an observer-to-environment map is constructed, and how a history of the environment can be obtained using the timekeeper's history as a reference of time. 

We will give a general formulation in arbitrary dimensions accompanied with examples in this section. Since the major example we have studied so far, the 1D gravitational path integral, will become too simple to be interesting after implementing a timekeeper, a way much richer example of 3D gravity incorporating AdS and closed universe and multiple observers/timekeepers will be constructed and studied in section \ref{sec:3D_gravity}. 

\subsection{The path integral relative to an observer}
We would like an observer to be a subsystem of the universe, who themselves are matter and can back react on the geometry. Therefore, the gravity action we consider is the same as \eqref{eq:L_grav_action}.
However, for convenience, we may separate the action of the observer $I_{\rm obs}$ from that of other matters $I_{\rm matter}$, and write it as 
\begin{align}\label{eq:L_grav_action_obs}
    I_{\rm grav} = \frac{1}{16\pi G_N} \int d^{d+1}x~\sqrt{-g}~(R-2\Lambda) + I_{\rm GH}+ I_{\rm matter} + I_{\rm obs}.
\end{align}
We consider the observer to be a $(p+1)$-D extended timelike object, where $p\leq d$. It can be a particle, a string, a brane, a union of these elements, etc. Let us use $\mathcal{M}$ to denote the geometry of the whole universe and $\mathcal{N}_{\rm obs}$ to denote the geometry of the observer's worldvolume. $\mathcal{N}_{\rm obs}$ is a subset of $\mathcal{M}$, i.e. $\mathcal{N}_{\rm obs} \subseteq \mathcal{M}$. The full path integral may be schematically written as 
\begin{align}\label{eq:full_PI}
    \sum_{\rm topology} \int {\mathcal{D}g} \left(\prod_{\mu=0}^d \prod_{i=0}^{p}\int\mathcal{D}X^\mu_i\right)\int\mathcal{D}\Phi \exp(iI_{\rm grav}), 
\end{align}
where $\prod_{\mu=0}^d \prod_{i=0}^{p}\int\mathcal{D}X^\mu_i$ denotes summing over the embedding of the observer's worldvolume $\mathcal{N}_{\rm obs}$ to the universe $\mathcal{M}$.
Note that this sum automatically includes different intrinsic geometries of the worldvolume $\mathcal{N}_{\rm obs}$. Also note that this path integral has no difference from \eqref{eq:fullPI}, since we are just summing over the configuration of the observer, who themselves are also ordinary matter. 

However, from the observer's point of view, they themselves are special. We may then define a path integral relative to the observer by putting restrictions using their specialness. Different sets of rules may be designed. 
The one we present in this subsection is a straightforward generalization of that proposed in \cite{AAIL25}.

Let us present some conditions which can be used to pick up favorable topologies. Consider a path integral with $n$ disconnected boundaries $A_1,A_2,\dots,A_n$. 
Let us first consider the case where the boundaries are divided into two subsets, which are regarded as $\Sigma_{\rm bra}$ and $\Sigma_{\rm ket}$, respectively.
\begin{itemize}
    \item {\bf Existence:} We may require that the boundary configurations imposed on $\Sigma_{\rm bra}$ and $\Sigma_{\rm ket}$ must have a nontrivial $p$-D overlap with the worldvolume of the observer $\mathcal{N}_{\rm obs}$. A boundary condition on $\Sigma_{\rm ket}$ with a trivial overlap can be interpreted as a state with no observer, which should not be encountered by the observer. 
    Therefore, we call this the {\it existence} condition. 
    \item {\bf Connectibility:} We may require that, 
    for any point on $\mathcal{N}_{\rm obs}$, there must exist a path on $\mathcal{N}_{\rm obs}$ connecting it to both $\Sigma_{\rm bra}\cap\mathcal{N}_{\rm obs}$ and $\Sigma_{\rm ket}\cap\mathcal{N}_{\rm obs}$. This implies that any piece of the observer in the future (past) should come from themselves in the past (future). We call this the {\it connectibility} condition. 
\end{itemize}
Let us then consider the case where the boundaries are divided into $2n$ subsets and regarded as $\Sigma_{\rm bra}^{(i)}$, $\Sigma_{\rm ket}^{(i)}$, where $1\leq i \leq n$ labels the replicas of the system. 
\begin{itemize}
    \item {\bf Replica Distinguishability:} We may require the existence condition and the connectibility condition to be satisfied within each replica. More precisely, both $\Sigma_{\rm bra}^{(i)}$ and $\Sigma_{\rm ket}^{(i)}$ should have nontrivial overlap with $\mathcal{N}_{\rm obs}^{(i)}$, and for any point on $\mathcal{N}_{\rm obs}^{(i)}$, there must exist a path on $\mathcal{N}_{\rm obs}^{(i)}$ connecting it to both $\Sigma_{\rm bra}^{(i)}\cap\mathcal{N}_{\rm obs}^{(i)}$ and $\Sigma_{\rm ket}^{(i)}\cap\mathcal{N}_{\rm obs}^{(i)}$. 
    This implies that different replicas of the observer should be distinguishable at the boundary.
    We call this the {\it replica distinguishability} condition. Note that we have not required $\mathcal{N}_{\rm obs}^{(i)} \neq \mathcal{N}_{\rm obs}^{(j)}$ for $i\neq j$. In fact, the case where all the boundaries are connected by a connected worldvolume satisfies this condition. 
    \item {\bf No replica collision:} We may further require $\mathcal{N}_{\rm obs}^{(i)} \cap \mathcal{N}_{\rm obs}^{(j)} = \emptyset$ for $i\neq j$, which implies the worldvolumes connecting each replica do not intersect. We call this the {\it no replica collision} condition. 
\end{itemize}
We can consider gravitational path integrals summing over topologies satisfying, for example, the first three conditions,  
or all the four conditions. 
Either choice will result in a violation of the permutation relation, and hence give a nontrivial Hilbert space dimension. Note that it is also possible to to consider other combinations of the conditions. 

The setup considered in \cite{AAIL25} is a special case under this rule, which is 2D gravitational path integrals with a 1D worldline observer satisfying all the four conditions above.

\subsubsection{Example: Observers in 1D gravity}
Let us see how this set of rules work in the 1D gravity setup considered in section \ref{sec:two_issues} as a simple example. As we have seen in section \ref{sec:NBwithEnsemble}, before identifying an observer, the full path integral with four boundaries $G_4(i,j,k,l)$ can be identified as several different objects such as $\overline{(\rho_{\rm NB})_{ij}(\rho_{\rm NB})_{kl}}$, $ \overline{(\rho_{{\rm NB},AB})_{ijkl}}$, $\overline{(|{\rm NB}\rangle_{AB})_{ij}(\langle {\rm NB}|_{AB})_{kl})}$ as shown in \eqref{eq:averaged_squared}, \eqref{eq:G4_2}, \eqref{eq:G4_3}. 

Let us now introduce a worldline observer described by the Nambu-Goto action. Such an observer action only results in a shift of the cosmological constant term, so the total action is essentially the same as that in section \ref{sec:two_issues}. However, the rules described above restricts topology. Let us consider the case where all of the four rules are applied.

Let us denote the no-boundary density matrix from the observer's perspective as $(\rho_{\rm NB|obs})$. $\overline{((\rho_{\rm NB|obs})_{ij}(\rho_{\rm NB|obs})_{kl}}$ involves two replicas, and the conditions lead to 
\begin{align}
    \overline{(\rho_{\rm NB|obs})_{ij}(\rho_{\rm NB|obs})_{kl}} = G_2(i,j)G_2(k,l),  
\end{align}
where the permutation invariance is broken. 
$\overline{((|{\rm NB|obs}\rangle_{AB})_{ij}(\langle {\rm NB|obs}|_{AB})_{kl})}$ also involves two replicas and is given by
\begin{align}
    \overline{((|{\rm NB|obs}\rangle_{AB})_{ij}(\langle {\rm NB|obs}|_{AB})_{kl})} = G_2(i,j)G_2(k,l), 
\end{align}
in a similar way. 

We can also consider an alternative observer choice described by two disjoint worldlines, which leads to 
\begin{align}
      \overline{(\rho_{{\rm NB|obs'},AB})_{ijkl}} = G_4(i,j,k,l),
\end{align}
where we have used obs$'$ because this is an observer choice different from the one above.

\subsection{The path integral relative to a timekeeper}
In the PW mechanism, after choosing an observer, one specifies a history of theirs to uplift the observer to a timekeeper. One then get a history of the environment via an observer-to-environment map, where the timekeeper's history is used as the reference of time. 

Now we would like to uplift the observer implemented in the gravitational path integral in the previous subsection to a timekeeper. Inspired by the PW mechanism, we would like to specify their history to make further restrictions on the gravitational path integral. 

A natural way to specify a history of an extended object is to fix the worldvolume geometry of it. In the previous subsection, we summed over all the configurations of $\mathcal{N}_{\rm obs}$ satisfying the boundary conditions and a set of topology-restricting conditions. Here, we fix $\mathcal{N}_{\rm obs}$ to $\mathcal{N}_{\rm tk}$, and sum over all the configurations of $\mathcal{M}$ compatible with an embedded $\mathcal{N}_{\rm tk}$. 

This path integral can be schematically written as 
\begin{align}\label{eq:Z}
    Z(\mathcal{N}_{\rm tk}) = \sum_{\rm topology} \int {\mathcal{D}g} \left(\prod_{\mu=0}^d \prod_{i=0}^{p}\int\mathcal{D}X^\mu_i \delta(\mathcal{N_{\rm obs},N_{\rm tk}})\right)~\int\mathcal{D}\Phi \exp(iI_{\rm grav})~, 
\end{align}
where $\delta(\mathcal{N_{\rm obs},N_{\rm tk}})$ is 1 when $\mathcal{N_{\rm obs}=N_{\rm tk}}$ and 0 otherwise. Note that we also need to impose the boundary conditions on temporal boundaries such as $\Sigma_{\rm bra}$ and $\Sigma_{\rm ket}$, etc, as we have done before, which is omitted in this equation. 

A crucial point is to notice that $Z(\mathcal{N}_{\rm tk})$ defines a map from the timekeeper's history $\mathcal{N}_{\rm tk}$ to a (class of) gravitational path integral.\footnote{Here, we have not specified the ``boundary" or ``junction" condition for the environment matter field $\Phi$ on $\mathcal{N}_{\rm tk}$, which serves as an undetermined degree of freedom. This is why we say ``a class of gravitational path integral".}
It is also worth noticing that this defines a non-gravitational theory on $\mathcal{N}_{\rm tk}$.
To see how and in what sense this map gives an environment's history more transparently, 
let us firstly consider a special case of a gravitational path integral relative to a timekeeper, which turns out to be the AdS/CFT correspondence \cite{Maldacena97,GKP98,Witten98}. This will also tell us how the gravitational path integral relative to a timekeeper \eqref{eq:Z} gives an observer dependent generalization of holography. 

\subsubsection{Example: the GKP-Witten formulation of AdS/CFT correspondence}
Consider a spacetime with a codimension-1 domain wall, where one side has a negative cosmological constant, and the other side has nothing.\footnote{This can be realized by starting from a negative cosmological constant $\Lambda$ and then taking $\Lambda\rightarrow -\infty$ the limit. See, e.g. \cite{BD11}. Such a domain wall is often called an end-of-the-world brane.} 
Let us regard this domain wall as the observer, and consider the special case where there is no temporal boundaries such as $\Sigma_{\rm bra}$ and $\Sigma_{\rm ket}$. 
If we fix the observer's worldvolume $\mathcal{N}_{\rm tk}$, which is infinitely large,\footnote{More precisely, we start from a finite-sized $\mathcal{N}_{\rm tk}$ and then take the large volume limit. Such $\mathcal{N}_{\rm tk}$ in general is not on-shell in the summing-over-$\mathcal{N}_{\rm obs}$ setup. However, in the case of Einstein graviy with $\Lambda=-1$, 
one way to make it on-shell in the summing-over-$\mathcal{N}_{\rm obs}$ setup is to consider a domain wall with tension $T=d-1$.} then the gravitational path integral $Z(\mathcal{N}_{\rm tk})$ is nothing but that appears in the GKP-Witten formulation \cite{GKP98,Witten98} of the AdS/CFT correspondence \cite{Maldacena97}. 

The AdS/CFT correspondence states that $Z(\mathcal{N}_{\rm tk})$ in this case is equivalent to a CFT defined on $\mathcal{N}_{\rm tk}$, which encodes the full dynamics of the bulk AdS, i.e. the environment in our language. Therefore, $Z(\mathcal{N}_{\rm tk})$ serves as an observer-to-environment map and describes a history of the environment using that of the observer as a reference of time, at least in this case which reduces to the AdS/CFT correspondence.

\subsubsection{Observer-dependent holography}

Considering that the gravitational path integral relative to a timekeeper \eqref{eq:Z} incorporates the GKP-Witten formulation of AdS/CFT as a special case, we can regard the gravitational path integral relative to a timekeeper as an observer-dependent generalization of holography. 

The holographic principle \cite{tHooft93,Susskind94} is usually stated in the following way: ``A quantum gravity living in a $(d+1)$-D spacetime is equivalent to a $d$-D non-gravitational quantum theory living on its boundary." Compared to this usual statement, the generalized version given by the gravitational path integral has the following feature: 
\begin{itemize}
    \item It is observer-dependent. This point manifests when there exist more than two extended objects in the spacetime. Taken one of them as the observer and further the timekeeper, the configurations for the other objects should be summed over just as what we do to the standard matter. Choosing another object as the observer/timekeeper will result in another \eqref{eq:Z}.
    \item The non-gravitational theory does not necessarily live on the boundary of the quantum gravity. This also happens in, for example, the DS/dS correspondence \cite{AKST04}, the static patch dS holography \cite{Susskind21,NV23} and the Cauchy slice holography \cite{AKW22}. 
    \item It is not necessarily a codimension-1 correspondence. This also happens in, for example, the wedge holography \cite{AKTW20} and the cone holography \cite{Miao21}.
    \item In fact, we may also consider a further generalization where the observer is not even necessarily an extended object but a general matter content, although it is not clear how to implement a gravitational path integral relative to such an observer, or how to uplift it to a timekeeper. 
\end{itemize}
These points will be made a little more concrete in the model constructed in section \ref{sec:3D_gravity}.

\subsubsection{Example: Timekeepers in 1D gravity}
As another example, let us see what happens if we introduce a timekeeper in the 1D gravity model. 
Let $\mathcal{N}_{\rm tk}$ be an interval of length $T$, the path integral relative to the timekeeper is nothing but a non-gravitational path integral on the interval, which describes a standard quantum mechanical time evolution. The implementation of a timekeeper is hence somewhat boring in the 1D case.

\section{A 3D gravity model}\label{sec:3D_gravity} 
In this section, we construct a 3D gravity model which incorporates all the elements discussed above, as well as the AdS/CFT correspondence \cite{Maldacena97,GKP98,Witten98}. This provides a more diverse playground than the 1D gravity model, our major example so far.

We start by writing down the the action of a 3D closed universe without observers or timekeepers. Then we proceed to identify an observer and discuss their uplift to a timekeeper. We will see that, in the end, the resulting theory with a timekeeper is what people usually called a gravitational path integral in an open universe, and includes a special well-understood case, the AdS/CFT correspondence. 
We hope this model serves as a starting point to study interesting quantities such as correlation functions and entanglement entropy, some of which will be discussed in upcoming works. 

\subsection{The action and its classical solutions}\label{sec:action_solution}

The theory we would like to consider is 3D Einstein gravity with a negative cosmological constant and an end-of-the-world brane $Q$ (i.e. a domain wall connecting the universe to nothing), given by the following (Lorentzian) action, 
\begin{align}\label{eq:ETW_action}
    I_{\rm grav} = \frac{1}{16\pi G_N} \int_{\CM} \sqrt{-g} (R+2) +\frac{1}{8\pi G_N} \int_Q \sqrt{-h} (K-T). 
\end{align}
The first term is the Einstein-Hilbert term and $\CM$ denotes the 3D manifold. $G_N$ is the 3D Newton constant, $g_{\mu\nu}$ is the metric and $R$ is the Ricci scalar of $\CM$. Note that we have set the cosmological constant to $\Lambda = -1$ for simplicity. The second term is the Gibbons-Hawking term on the end-of-the-world brane $Q$. $h_{ab}$ denotes the intrinsic metric, $K_{ab}$ is the extrinsic curvature, and we take the normal vector to point outwards. $T$ is the tension of the end-of-the-world brane for which we consider $T>1$. 

Later we would like to consider the gravitational path integral summing over all the configurations of $\mathcal{M}$ and $Q$ with respect to the action $I_{\rm grav}$. Before that, let us have a look at its saddle points, i.e. solutions of the equation of motion. Because we will sum over all possible worldvolume of $Q$ later, the proper boundary condition to impose on the saddle points of $I_{\rm grav}$ is the Neumann boundary condition. 
\begin{align}\label{eq:NBC}
    K_{ab} - Kh_{ab} + Th_{ab} = 0. 
\end{align}
The choice of the tension range $T>1$ turns out to be crucial since it admits smooth solutions without the necessity to specify boundary data on any asymptotic regions, as we will see soon. 

Let us start by looking at a Euclidean saddle point.  
The most simple one is a 3D Poincar\'e ball with a finite cutoff which is parameterized by $\phi\in[0,2\pi)$, $\theta \in [-\pi/2, \pi/2]$ and $\eta \in [0,\eta_*]$. The metric is 
\begin{align}
    ds^2 = d\eta^2 + \sinh^2 \eta (d\theta^2 + \cos^2 \theta d\phi^2), 
\end{align}
and 
\begin{align}
    \eta_* = {\rm arccoth}(T), 
\end{align}
is the location of the end-of-the-world brane $Q$. The profile of $Q$ is a 2-sphere. 
This solution describes a bubble which is locally AdS$_3$ anywhere except for the vicinity of $Q$.
However, it does not have an asymptotic boundary and hence is not asymptotically AdS. Let us denote it as $\CM = \CM_{E,1}$. A simple computation leads to the on-shell Euclidean action on it, 
\begin{align}
    I_{E, \rm grav}[\CM_{E,1}] = \frac{1}{2G_N} {\rm arccoth}(T). 
\end{align}
Performing the Wick rotation $\theta \rightarrow it$, we obtain a Lorentzian solution $\mathcal{M}_{L,1}$ whose metric is 
\begin{align}\label{eq:eta_t_phi}
    ds^2 = d\eta^2 + \sinh^2 \eta (-dt^2 + \cosh^2 t d\phi^2), 
\end{align}
where the brane profile $Q$ now is a global dS$_2$. Note that the $(t,\eta,\phi)$-coordinate used here does not cover the entire 3D spacetime. Especially, the null surface $\eta = 0$ turns out to be an event horizon from the brane's perspective. The region beyond this event horizon can be described by the $(\tau,\rho,\phi)$-coordinate obtained from 
\begin{align}
    \cosh~t = \frac{\sinh~\rho}{\sinh ~\eta},~~\sinh~t = \frac{\tan~\tau}{\tanh~\eta}, 
\end{align}
with which the metric is 
\begin{align}
    ds^2 = d\rho^2 -\cosh^2\rho d\tau^2 + \sinh^2 \rho d\phi^2,
\end{align}
and the brane profile is given by 
\begin{align}
    T^2 = \frac{\cosh^2 \rho}{\sinh^2 \rho - \tan^2 \tau}.
\end{align}
The location of the event horizon is now $\rho = {\rm arcsinh} \tan \tau$.
Note that the brane only exists in $-\pi/2 \leq \tau \leq \pi/2$ in this coordinate. The $\rho < {\rm arcsinh} \tan \tau$ can be regarded the ``black hole interior" in this geometry. 

Each spatial slice $\tau = \rm const.$ in this geometry is a finite-volumed disk, the size of which first shrinks and then expands as $\tau$ goes from $-\tau/2$ to $\tau/2$. 
Such a configuration can be regarded as a Coleman-de Luccia instanton describing a tunneling process from nothing to AdS, similar to that considered in \cite{Maldacena10}.

As a manifold, a disk is neither closed nor open. Therefore, it is neither a closed universe or an open universe in the usual sense. However, since both $\mathcal{M}$ and $Q$ are dynamically determined and no Dirichlet type constraints are imposed in this theory, this is essentially a closed universe theory. In fact, there are two simple ways to relate this theory to a standard closed universe theory. The first is to consider an alternative theory including a domain wall connecting two AdS Einstein gravity: 
\begin{align}
    I_{\rm grav}' = \frac{1}{16\pi G_N} \int_{\CM} \sqrt{-g} (R+2) +\frac{1}{8\pi G_N} \int_{\rm domain~wall} \sqrt{-h} (K_L-K_R-2T), 
\end{align} 
where $K_{L(R)}$ is the extrinsic curvature on the left (right) side of the domain wall and the normal vector points from left to right. A solution of this theory can be constructed from a solution of the end-of-the-world brane theory \eqref{eq:ETW_action} in the following way. Take a solution of \eqref{eq:ETW_action} e.g. $\CM_{L,1}$, prepare two copies of it, and glue them along their boundaries. The resulting geometry is a solution of the domain wall model and it is a closed manifold. Similarly, we can see that the theory \eqref{eq:ETW_action} we focus on is equivalent to the $\mathbb{Z}_2$-symmetric sector of this standard closed universe theory. The second way is to realize that an end-of-the-world brane can be obtained from performing dimensional reduction to a higher-dimensional closed manifold (See e.g. \cite{SS23}). For example, shrinking an $S^1$ direction on $S^2$ can result in an interval, which is a 1D manifold with a 0D end-of-the-world brane and encodes the dynamics in $S^2$.  

Given the reasoning above, we will simply call solutions like $\mathcal{M}_{E,1}$ and $\mathcal{M}_{L,1}$ closed universe solutions in the following. 

Before moving on to the gravitational path integral, we would like to note that this theory also admit asymptotically AdS solutions. So far we solved the classical equation of motion under the condition that the asymptotic boundary is an empty set. But we can also consider a non-trivial asymptotic boundary, e.g. a $S^2$. Then, of course, the Euclidean AdS$_3$ is a solution, where the brane $Q$ has a vanishing worldvolume. Besides, the union between the Euclidean AdS$_3$ and a $\mathcal{M}_{E,1}$ is also a solution for this case.

\subsection{Path integral, observer, timekeeper and holography}
The full gravitational path integral sums over all the possible worldvolume of $Q$ and $\mathcal{M}$ compatible with it. This suffers from both the problem of time and the problem of dimension. 

By identifying the brane $Q$ as the observer, one can use the conditions discussed in section \ref{sec:observer_timekeeper} to discard certain topologies to break the permutation symmetry of the path integral with multiple boundaries, and thus avoid the problem of dimension. 

Let us focus on the case where we fix the the worldvolume of $Q$ to $\mathcal{N}_{\rm tk}$ to uplift them to a timekeeper, and only sum over all the $\mathcal{M}$ satisfying $\partial \mathcal{M} = \mathcal{N}_{\rm tk}$. The path integral can be schematically written as 
\begin{align}
    Z_{|Q}(\mathcal{N}_{\rm tk})
    = \int_{} Dg_{\mu\nu} \exp\left(\frac{i}{16\pi G_N} \int_{\CM} \sqrt{-g} (R+2) +\frac{i}{8\pi G_N} \int_{\mathcal{N}_{\rm tk}} \sqrt{-h} (K-T)\right),  
\end{align}
where $|Q$ means that this path integral is defined relative to regarding $Q$ as the timekeeper. 
We are now doing a path integral on a restricted configuration space, where the restricted saddle points with respect to the action \eqref{eq:ETW_action} are obtained by solving the equation of motion with the Dirichlet boundary condition 
\begin{align}\label{eq:DBC}
    \delta{h_{ab}}= 0
\end{align}
on $Q$. These saddle points serve as constrained instantons in the original path integral problem without a timekeeper, and are not solutions to the Neumann boundary condition in general. 

This path integral defines a non-gravitational theory on $Q=\mathcal{N}_{\rm tk}$. 
In fact, it has been conjectured that such a gravitational path integral is effectively described by a $TT$ deformed CFT defined on $Q=\mathcal{N}_{\rm tk}$ \cite{MMV16}. The specific case where $\mathcal{N}_{\rm tk}$ is a sphere is studied in \cite{CDS19}, and see \cite{HKST18} for higher dimensional cases. 

We may also consider the following very special case. Fix $Q=\mathcal{N}_{\rm tk}$ such that the metric on it is 
\begin{align}
    ds^2 = \sinh^2\eta_{\infty} \left(d\theta^2 + \cos^2\theta d\phi^2\right), 
\end{align}
and send $\eta_\infty \rightarrow \infty$. Again, such a $Q=\mathcal{N}_{\rm tk}$ does not need to be a solution to the Neumann boundary condition \eqref{eq:NBC}, but if we want it to be, this can be achieved by taking the $T\rightarrow1$ limit. For such a $Q=\mathcal{N}_{\rm tk}$, the path integral is known to be described by a CFT defined on manifolds which are conformally equivalent to $Q=\mathcal{N}_{\rm tk}$. This is nothing but the GKP-Witten formulation \cite{GKP98,Witten98} of the AdS/CFT correspondence \cite{Maldacena97}.

\subsection{Another candidate observer and observer dependence}\label{sec:entanglement} 
Let us introduce another matter content which may be also regarded as an observer. We add a point particle $P$ with mass $m$
\begin{align}
    I_{\rm grav} = \frac{1}{16\pi G_N} \int_{\CM} \sqrt{-g} (R+2) +\frac{1}{8\pi G_N} \int_Q \sqrt{-h} (K-T) - m\int_P\sqrt{-\gamma}. 
\end{align}
Let us present a classical solution of this action. When $m\ll 1/G_N$, $P$ can be regarded as a probe particle which does not backreact on the geometry. Thus, the Lorentian configuration $\mathcal{M}_{L,1}$ discussed in the previous section accompanied with a world line sitting at $\rho =0$ serves as a classical solution of this new theory. In this solution, the brane $Q$ sits outside the event horizon, while $P$ is always inside the event horizon. In this sense, this configuration can be regarded as a toy model of a black hole, where one observer stays outside and the other sits inside. 

A full path integral sums over all possible worldvolume $Q$, worldline $P$, and $\mathcal{M}$ compatible with them. We can at least consider the following three types of gravitational path integral. The first $Z_{|P}$ regards $P$ as the timekeeper and sum over all possible configurations of $Q$ and $\mathcal{M}$. The second $Z_{|Q}$ takes $Q$ as the timekeeper and sum over all possible configurations of $P$ and $\mathcal{M}$. The third $Z_{|P\cup Q}$ takes $P\cup Q$ as the timekeeper and only sum over $\mathcal{M}$ compatible with them. This contrasts an observer dependence of the gravitational path integral relative to an observer/timekeeper.

\section{Conclusions and discussions}\label{sec:discussion}
In this paper, we first compare the origin of the problem of time and that of the problem of dimension in the path integral quantization of gravity in section \ref{sec:two_issues}. The former arises from summing over metrics too nicely while the latter arises from summing over topologies too nicely. It follows immediately that introducing a generic unbalanced weight in summing over metrics (topologies) would avoid the problem of time (dimension), and the question is to find a physically sensible way to do so. 

We then revisit the PW mechanism \cite{PW83,Wootters84} in section \ref{sec:PW}, which is one of the earliest proposal tackling with the two problems, but in a purely quantum mechanical and non-gravitational setup. In the PW mechanism, one starts from a single static state of the universe $\ket{\Psi(t)}_{\rm uni}$ and then identify a subsystem of the universe as an observer. The observer will be able to see a nontrivial Hilbert space dimension thanks to entanglement structure between themselves and their complement, i.e. the environment. In order to obtain a nontrivial time evolution from the observer's point of view, one needs to specify a history of the observer and uplift them to a timekeeper. An environment's history is then obtained by performing an observer-to-environment map constructed from $\ket{\Psi(t)}_{\rm uni}$ to the timekeeper's history. 

Inspired by the PW mechanism, we present an implementation of observers and timekeepers in the gravitational path integral in section \ref{sec:observer_timekeeper}. We first identify an extended object in the spacetime, which itself is matter, as the observer, and present a set of conditions one may use to choose which topologies to sum over. This part is a straightforward generalization of the rules introduced in \cite{AAIL25}. We then describe how to specify a history of the observer to uplift them to a timekeeper, by fixing the intrinsic geometry of their worldvolume in the gravitational path integral. Such a process introduces a map from the timekeeper's history to a (class of) gravitational path integral, which is regarded as the environment's history using the timekeeper's history as a reference of time. We discuss its relation with the AdS/CFT correspondence, where the asymptotic boundary of AdS is regarded as the timekeeper, and the environment's history can be described by a CFT on the asymptotic boundary. We have also discussed how the gravitational path integral relative to a timekeeper serves as a generalization of holography. This notion of holography is manifestly observer dependent, and contains different information from each other. 

We have also seen how the implementation works in 1D gravity in section \ref{sec:observer_timekeeper}, but it turns out to be too simple to be interesting when a timekeeper is involved. Therefore, in section \ref{sec:3D_gravity}, we have constructed a 3D model as an example incorporating all the elements discussed in the previous sections, including closed universes, AdS (which is an open universe), multiple observers and timekeepers. We expect this model serves as a starting point for further studies of gravitational path integrals involving multiple observers and timekeepers. 

Throughout the way, the relation between the PW mechanism and the relative formulation of gravitational path integral has been made clear. 
Since the PW mechanism plays an important role in the study of quantum reference frame \cite{AS67,BRS07,GS08,GCB17,VHGC18,HSL19}, this connection, although being qualitative, is expected to help us use well-developed results in quantum reference frames to study relational gravitational path integrals. On the other hand, observations from the gravitational path integral may also provide interesting inputs for the study of quantum reference frame. We hope this paper provides a starting point to bring the two streams together.  

We outline several future directions in this spirit, which we will revisit and address in upcoming work.

\subsubsection*{Observer dependence and switching map} 
The gravitational path integral relative to an observer or a timekeeper is observer-dependent. In particular, the gravitational path integral relative to a timekeeper also serves as an observer-dependent generalization of holography. Given different physics observed by different observers/timekeepers, it is natural to ask which part of physical observations they can share with each other. 

This question is relatively better-understood in the PW mechanism (or more generally the quantum reference frame). In this context, different choices of the observer give arises to apparent different physics, but gauge-invariant information can be shared among the observers via a ``reference frame switching map" \cite{GCB17,VHGC18}. This structure is very similar to that of a quantum error correction code \cite{Rothlin24,CCHM24}. On the other hand, it is also known that holography can act as quantum error correction codes \cite{ADH14,PYHP15,Harlow16,AEHPV22}. Therefore, taking existing results in quantum references frames as hints, 
it should be promising to understand the observer-dependence of gravity and holography in better details and develop a switching map between different observers. 

\subsubsection*{Quantum reference frame with 1D Hilbert space}
It would also be interesting to study the quantum reference frame itself with some inputs from the gravitational path integral. For example, while the gravitational path integral predicts the fundamental Hilbert space to be 1D in a closed universe before specifying an observer, such setup has not been considered in quantum reference frame literature. It should be interesting to see what consequences this strong input can give in a purely (non-gravitational) quantum mechanical setup.

\subsubsection*{Inter-observer entanglement entropy}
In a setup involving two (candidate) observers, e.g. the setup in section \ref{sec:3D_gravity} involving a particle observer and a brane observer, it is possible to compute the entanglement entropy between the two observers via the replica trick \cite{CC04, LM13, PSSY19,AHMST19}. It would be interesting to study the differences in the entanglement structure among different treatments of the observers. In particular, when both observers are treated as timekeepers and the union of them coincides the asymptotic boundary of an AdS spacetime, the entanglement entropy should reduce to the area of an extremal surface, as described by the 
Ryu-Takayanagi formula \cite{RT06,RT06b}.

\section*{Acknowledgements}
I am grateful to Jordan Cotler, Kanato Goto, Daniel Jafferis, Ho-Tat Lam, Andy Strominger, Tadashi Takayanagi, Diandian Wang, Jinzhao Wang, Zhencheng Wang and Mengyang Zhang for useful discussions and
comments. I would also like to thank useful
comments from participants of Celestial Holography Workshop 2025, where a part of this
work was presented. I am supported by the Society of Fellows at Harvard University.

\appendix

\section{Observer-environment entanglement in 1D gravity} \label{app:observerEE_1D}

We have seen in section \ref{sec:PW} that in the PW mechanism, the Hilbert space dimension an observer can see is characterized by its entanglement with the environment. Let us study this entanglement structure in the 1D gravity coupled to scalar fields model studied in section \ref{sec:two_issues}.

We consider the no-boundary wave function $\ket{\rm NB}_{AB}$ in \eqref{eq:NB_AB}. Note that at this stage, we have not yet included contributions from summing over topologies, and $\ket{\rm NB}_{AB}$ is regarded as a pure state in $\mathcal{H}_{{\rm con},A} \otimes \mathcal{H}_{{\rm con},B}$. It satisfies two independent Hamiltonian constraints as shown in \eqref{eq:two_constraints}. Let us write down the state more explicitly. 

When the number of species $M=1$, the Hamiltonian can be diagonalized as 
\begin{align}
    H = \sum_{n=0}^{\infty} E^{(n)} |n\rangle\langle n|, 
\end{align}
where 
\begin{align}
    E^{(n)} = \left(n+\frac{1}{2}\right)m + \frac{\Lambda}{8\pi G_N}, 
\end{align}
Accordingly, $\ket{\rm NB}_{AB}$ is not null only when 
\begin{align}
    -\frac{\Lambda}{8\pi m G_N}-\frac{1}{2} = 0,1,2,..., 
\end{align}
and turns out to be 
\begin{align}
    \ket{\rm NB}_{AB} = \left|-\frac{\Lambda}{8\pi m G_N}-\frac{1}{2}\right\rangle_A\left|-\frac{\Lambda}{8\pi m G_N}-\frac{1}{2}\right\rangle_B, 
\end{align}
which is merely a product state between the two sides. 

For general species number $M$, the Hamiltonian can be diagonalized as 
\begin{align}
    H = \sum_{n_1=0}^{\infty}\cdots\sum_{n_M=0}^{\infty} \left(E^{(n_1)}+\cdots+E^{(n_M)} +  \frac{\Lambda}{8\pi G_N}\right) |n_1\dots n_M\rangle\langle n_1\dots n_M|.
\end{align}
Accordingly, 
\begin{align}
\ket{\rm NB}_{AB} = \sum_{(n_1,n_2,\dots, n_M)~{\text s.t. }\sum n{_i }= N } |n_1n_2... n_M\rangle_A|n_1n_2... n_M\rangle_B,~~~{\rm where}~N = -\frac{\Lambda}{8\pi m G_N} - \frac{M}{2}. 
\end{align}
The number of members appearing in this sum is a standard stars and bars combinatorics, 
\begin{align}
     \#\left({(n_1,n_2,\dots, n_M)~{\text s.t. }\sum n{_i }= N }\right) = \binom{M+N-1}{M-1}. 
\end{align}

Let us regard the first $J$ species of scalar fields living in the $A$ side of the universe as the ``observer" and investigate its entanglement with the other scalar fields viewed as the ``environment". We study two quantities, the rank and the entanglement entropy of the observer's reduced density matrix, both of which characterizes the effective dimension of the Hilbert space visible to the observer. 

For the rank, the problem is reduced to computing the number of cases in the following setup. Consider $J$ boxes with labels and $K$ balls without labels, distribute the $K$ balls in them, and sum over $K = 0, 1,2,...,N$. The resulting number of cases is 
\begin{align}
    \sum_{K=0}^N \binom{J+K-1}{J-1} = \sum_{K=0}^{N} \binom{J+K-1}{K} = \binom{J+N}{J}
\end{align}
where we have used the hockey-stick identity in the last line. Keeping $J$ finite and taking the semiclassical limit $G_N\rightarrow0$, $\binom{J+N}{J} \sim N^J$. Therefore, the corresponding entropy is $\log \binom{J+N}{J} \approx J \log N$, which is $\Theta(J\log (1/G_N))$. On the other hand, if we take $G_N\rightarrow0$ while keeping $J/N$ finite, $\log \binom{J+N}{J} = \Theta(1/G_N)$. In either case, the observer will be able to see an infinite dimensional Hilbert space at the semiclassical limit $G_N\rightarrow0$. 

For the entanglement entropy, the problem is reduced to the following. Consider $M$ boxes with labels and $N$ balls without label and distribute the $N$ balls into $M$ boxes. Among all $\binom{M+N-1}{N}$ cases with the same probability, we want to compute the Shannon entropy of the reduced probability distribution for the first $J$ boxes. There are $\binom{K+J-1}{K}$ ways to distribute $K$ balls in the first $J$ boxes, 
and for each case, there are $\binom{(N-K)+(M-J)-1}{N-K}$ ways to distribute $N-K$ balls in the last $(M-J)$ boxes. Therefore, the probability for the configuration $(n_1,n_2,...,n_J)$ to appear for the first $J$ boxes is 
\begin{align}
    P(n_1,n_2,...,n_J) = \frac{\binom{(N-\sum_{i=1}^J n_i)+(M-J)-1}{N-\sum_{i=1}^J n_i}}{\binom{N+M-1}{N}}. 
\end{align}
This itself is complicated, but we can investigate its behavior at certain limits. Since what we are doing is nothing but computing entropy for a $M$-oscillator system in the microcanonical ensemble with energy $N+M/2$. Therefore, we expect the volume law of entropy to hold. 

Assuming $N,M\gg 1$, and $N/M = \Omega(1)$, we have 
\begin{align}
    S_{\rm total} &= \log \binom{N+M -1}{N} \nonumber \\
    &\approx (N+M) \log (N+M) - (N+M) -N \log N +N - M\log M + M
    \\
    &\approx (N + M ) \log (N+M) - N \log N - M \log M. 
\end{align}
Accordingly, the entropy of the observer is 
\begin{align}
    S_{\rm obs} &= \frac{J}{M} \left[ (N + M ) \log (N+M) - N \log N - M \log M \right] \nonumber \\
    &= J \left[  \log \left(1 + \frac{N}{M} \right) - \frac{N}{M} \log \left(1 + \frac{M}{N} \right)\right]. 
\end{align}
Here $J$ is a finite fraction of $M$ and hence $\Theta(1/G_N)$, while the $[\dots]$ term is $\Theta(1)$. 

As a summary, in the 1D model studied in section \ref{sec:two_issues}, we can identify $J$ species of the scalar fields as the observer and the complement as the environment. 
In this setup, if $J$ itself is a non-vanishing fraction of $N$, the effective entropy of their Hilbert space is $\Theta(1/G_N)$ at the semiclassical limit $G_N\rightarrow0$. If $J$ itself is finite, then the effective entropy of their Hilbert space is $\Theta(\log(1/G_N))$.

\bibliographystyle{jhep}
\bibliography{PW_Holography}

\end{document}